\documentclass[onecolumn,authoryear]{els-mrw}

\newcommand{\MJysr}{{\rm MJy\,sr^{-1}}}

\newcommand{\Mpc}{{\rm Mpc}}
\newcommand{\GHz}{{\rm GHz}}

\newcommand{\expf}[1]{{{\rm e}^{#1}}}
\newcommand{\Jbb}{\mathcal{J}}
\newcommand{\JDC}{\mathcal{J}_{\rm DC}}

\newcommand{\Tz}{{T_{z}}}

\newcommand{\TCMB}{T_{\rm CMB}}

\newcommand{\zh}{{z_{\rm h}}}
\newcommand{\zmu}{{z_{\mu}}}
\newcommand{\zmudc}{{z_{\rm dc}}}

\newcommand{\nbb}{{n_{\rm bb}}}

\newcommand{\nS}{n_{\rm S}}

\newcommand{\nrun}{n_{\rm run}}

\newcommand{\Planck}{{\it Planck}\xspace}

\newcommand{\x}{{\hat{x}}}

\newcommand{\kD}{k_{\rm D}}

\newcommand{\xe}{x_{\rm e}}

\newcommand{\xc}{x_{\rm c}}

\newcommand{\id}{{\,\rm d}}

\newcommand{\beq}{\begin{equation}}   %

\newcommand{\eeq}{\end{equation}}   %

\newcommand{\beqa}{\begin{eqnarray}}   %

\newcommand{\eeqa}{\end{eqnarray}}   %

\newcommand{\bspl}{\begin{split}}

\newcommand{\espl}{\end{split}}

\newcommand{\beal}[1]{\begin{align} #1 \end{align}}
\newcommand{\bsub}{\begin{subequations}}
\newcommand{\esub}{\end{subequations}}

\newcommand{\bmulti}{\begin{multline}}   %

\newcommand{\beqm}{\begin{mathletters}}   %

\newcommand{\eeqm}{\end{mathletters}}   %

\newcommand{\kB}{k_{\rm B}}

\newcommand{\me}{m_{\rm e}}

\newcommand{\Ne}{N_{\rm e}}

\newcommand{\Te}{T_{\rm e}}

\newcommand{\Tg}{T_{\gamma}}

\newcommand{\The}{\theta_{\rm e}}

\newcommand{\sigT}{\sigma_{\rm T}}

\newcommand{\pot}[2]{#1 \times 10^{#2}}

\newcommand{\Thz}{\theta_{z}}
\newcommand{\Yp}{Y_{\rm p}}

\newcommand{\zeq}{z_{\rm eq}}

\usepackage{amsmath,amssymb,amsfonts,amsthm,makeidx,graphicx}
\usepackage{txfonts}
\usepackage{helvet}
\usepackage{wrapfig}
\usepackage[resetlabels,labeled]{multibib}
\usepackage{lipsum}
\usepackage{xspace, natbib, color}
\usepackage{comment}

\newcommand{\listindentation}{5mm}

\newcommand{\aap}{A\&A}

\newcommand{\apj}{ApJ}
\newcommand{\apjl}{ApJL}
\newcommand{\apjs}{ApJS}

\newcommand{\jcap}{JCAP}

\newcommand{\mnras}{MNRAS}
\newcommand{\nat}{Nature}

\newcommand{\prd}{Phys. Rev. D}
\newcommand{\araa}{Ann. Rev. A.\&A.}
\newcommand{\apss}{ApSS}
\newcommand{\baas}{BAAS}

\newcommand{\COBE}{{\it COBE}\xspace}
\newcommand{\COBEF}{{\it COBE/FIRAS}\xspace}
\newcommand{\PIXIE}{{\it PIXIE}\xspace}
\newcommand{\PRISM}{{\it PRISM}\xspace}
\newcommand{\PRISTINE}{{\it PRISINE}\xspace}
\newcommand{\FOSSIL}{{\it FOSSIL}\xspace}
\newcommand{\WMAP}{{\it WMAP}\xspace}
\newcommand{\Litebird}{{\it Litebird}\xspace}

\newcommand{\eV}{{\rm eV}}
\newcommand{\keV}{{\rm keV}}

\newcommand{\Kel}{{\rm K}}

\newcommand{\cm}{{\rm cm}}
\newcommand{\km}{{\rm km}}
\newcommand{\Hz}{{\rm Hz}}
\newcommand{\ergs}{{\rm ergs}}
\newcommand{\sr}{{\rm sr}}

\newcommand{\Omegab}{\Omega_{\rm b}}
\newcommand{\Omegac}{\Omega_{\rm c}}
\newcommand{\OmegaL}{\Omega_{\Lambda}}
\newcommand{\taure}{\tau_{\rm re}}
\newcommand{\AS}{A_{\rm S}}
\newcommand{\Neff}{N_{\rm eff}}
\newcommand{\Nbary}{N_{\rm b}}

\begin{document}

\chapter{The Cosmic Microwave Background: Spectral Distortions}\label{chap1}

\author[1]{\vspace{2mm}Jens Chluba\vspace{2mm}}
\address[1]{
\orgname{Jodrell Bank Centre for Astrophysics, School of Physics and Astronomy}, 
\orgaddress{The University of Manchester, Manchester M13 9PL, U.K.}}

\articletag{Chapter Article tagline: update of previous edition, reprint.}

\maketitle

\begin{glossary}[Glossary]
\vspace{-2mm}
\term{Compton equilibrium temperature}  -- Temperature that the electrons reach in a given (distorted) photon field through Compton scattering.
\\
\term{Comptonization}  -- Evolution of the photon distribution under repeated Compton scattering.
\\
\term{Cosmological recombination era} -- Phase in the history of the Universe during which hydrogen and helium became neutral.
\\
\term{Cosmological recombination radiation} -- Spectral distortion imprinted by the recombining hydrogen and helium atoms at redshift $\simeq 10^3$.
\\
\term{Distortion visibility function} -- Function that describes which fraction of the injected energy is still visible as a spectral distortion today.
\\
\term{Last scattering surface} -- Imaginary redshift surface at which photons scatter for the last time with the free electrons.
\\
\term{Spectral distortion} -- Departure of the photon distribution from that of a perfect blackbody
\\
\term{Thermal SZ effect} -- Spectral distortion to the CMB in the direction of galaxy clusters due to Compton scattering by thermal electrons
\\
\end{glossary}

\begin{glossary}[Nomenclature and Abbreviations]
\vspace{-2mm}
\begin{tabular}{@{}lp{34pc}@{}}
BBN & Big Bang Nucleosynthesis
\\
BH & black hole
\\
CDM & Cold Dark Matter model 
\\
CMB & Cosmic Microwave Background
\\
GR & General Relativity
\\
$\Lambda$CDM & $\Lambda$ Cold Dark Matter model 
\\
LTE & Local Thermodynamic Equilibrium
\\
$N_{\rm eff}$ & Effective number of neutrino species
\\
PBH & primordial black hole
\\
SD & Spectral Distortions
\\
SKA & Square Kilometer Array
\\
SM & Standard Model of Particle Physics
\\
SZ & Sunyaev-Zeldovich effect
\\
$Y_{\rm p}$ & Primordial mass fraction of $^{4}{\rm He}$
\\
\end{tabular}
\end{glossary}

\begin{glossary}[CMB missions, spectrometers and experiments]
\vspace{-2mm}
\begin{tabular}{@{}lp{34pc}@{}}
\COBE & Cosmic Background Explorer launched by NASA in 1989
\\
\WMAP & Wilkinson Microwave Anisotropy Probe launched by NASA in 2001
\\
\Planck & L-class CMB mission launched by ESA in 2009
\\
\Litebird & Next generation CMB mission planned by JAXA
\\
{\it PICO} & Ambitious future CMB mission concept in the US
\\[1mm]
\PIXIE & CMB spectrometer concept first discussed in 2011
\\
\PRISM & CMB imager and spectrometer concept proposed to ESA for the L2 and L3 call in 2013
\\
\FOSSIL & CMB spectrometer concept proposed to ESA for the M7 call in 2022
\\
Voyage 2050 & Long-term programmatic planning of ESA's future space program
\\[1mm]
ACT & Atacama Cosmology Telescope
\\
SPT & South Pole Telescope
\\
SO & The Simons Observatory
\\
CMB-S4 & Stage IV CMB experiment
\\
TMS & Tenerife Microwave Spectrometer 
\\
BISOU & CMB spectrometer balloon supported by CNES

\end{tabular}

\end{glossary}

\newpage

\begin{abstract}[Abstract]
The cosmic microwave background (CMB) traveled the cosmos long before it reached our telescopes today. Consequently, it is one of the best probes of fundamental processes in the early Universe that we could hope to observe.
The cosmological information is encoded in two distinct ways. First, we can investigate how the CMB photons in one sky-direction are distributed across energy by focusing on information carried by the {\it CMB frequency spectrum}. Second, we can compare the flux of CMB photons that we receive from different directions, this time at a fixed frequency, to study the {\it CMB anisotropies}.
In the past six decades since the serendipitous discovery of the CMB in 1965, cosmologists have advanced both frontiers in terms of theory and observation. 
In this chapter, I will give a broad-brush overview about how the CMB spectrum forms and evolves throughout cosmic history, mentioning CMB anisotropies only on the side. I will attempt to highlight some of the key theoretical ingredients that allowed us to establish the detailed picture of the Universe we have today.
With this, I hope to convince you that, beyond the impressive past successes, the CMB still holds many treasures for us, and will keep generations of scientists busy for the decades to come.
\end{abstract}

\begin{keywords}
Cosmic microwave background, CMB spectral distortions, early universe physics, cosmological parameters.
\end{keywords}

\section*{Key points and learning objectives}
\begin{itemize}
\setlength{\itemindent}{\listindentation}
    \item Grasp why the CMB has been so instrumental to cosmology
    \item Learn about the creation of CMB spectral distortions in the early Universe
    \item Build a perspective on the future of CMB distortion science
\end{itemize}

\vspace{-3mm}
\section{Introduction and motivation}
To understand the properties of the Universe we live in, we must study signals from the earliest phases of cosmic history. The oldest light we can directly observe is that of the cosmic microwave background (CMB). Starting with its discovery in\footnote{We quote the year of the related paper \citep{Penzias1965} although the measurements were carried out in 1964.} 1965 by Penzias and Wilson\footnote{Nobel Prize in Physics in 1978} to the most recent all-sky temperature maps obtained with the \Planck surveyor, studies of the CMB have undoubtedly been instrumental to establishing the Cosmological Standard Model, known as the $\Lambda$ Cold Dark Matter model ($\Lambda$CDM). This model is based on General Relativity (GR) and the Standard Model of Particle Physics (SM), with extra ingredients from cosmic {\it Inflation} physics to set the initial conditions. By studying the CMB, we can learn about all these ingredients, giving answers to big questions like: {\it What is the Universe made of?} {\it What seeded the structures in the Universe?} {\it How did the structures evolve in time?} {\it Is there more to the Universe than meets the eye?}

The past six decades of CMB research have helped to transform the field of Cosmology into a precise scientific discipline, describing a wealth of very accurate measurements with one theoretical model. The {\it seven} key cosmological parameters are: 
\begin{itemize}
\setlength{\itemindent}{\listindentation}
\vspace{1mm}
\item the average CMB temperature, $T_0=(2.7255\pm 0.0006)\Kel$
\item the Hubble parameter, $H_0=(67.66 \pm 0.42)\,\km \sec^{-1}\Mpc^{-1}$
\item the energy density contribution from baryons, $\Omegab h^2=0.02242 \pm 0.00014$
\item the energy density contribution from cold dark matter, $\Omegac h^2=0.11933 \pm 0.00091$
\item the reionization optical depth, $\taure=0.0561 \pm 0.0071$
\item the amplitude of scalar perturbations, $\AS$, usually given in the form $\ln(10^{10} \AS)=3.047 \pm 0.014$
\item the spectral index of scalar perturbations, $\nS=0.9665 \pm 0.0038$,
\vspace{1mm}
\end{itemize}
where we used $h=H_0/[100 \,\km \sec^{-1}\Mpc^{-1}]$. To derive these constraints, we assumed the standard BBN model, which predicts the effective number of relativistic degrees of freedom (from three species of SM neutrinos), $\Neff=3.046$, and the helium mass fraction, $\Yp=0.2467$. We furthermore relied on standard atomic physics and radiative transfer to compute the standard ionization history of the Universe around the time of recombination, some 380,000 years after the Big Bang, and during reionization at redshift $z\simeq 10$. And finally, we assumed a flat cosmology, with a cosmological constant constituting the remaining $\simeq 70\%$ of the energy density, $\OmegaL=1-\Omegab-\Omegac\approx 0.6889 \pm 0.0056$.

Postponing questions about what all these parameters actually mean, at this point, the reader should be {\it amazed}: A theoretical model -- $\Lambda$CDM -- with {\it only} seven free parameters allow us to describe our current cosmological observations. Or put another way, we measured the CMB in millions of patches and can compress all that information into just seven numbers. Furthermore, we know most of these parameters to percent-level precision or better. This clearly is an outstanding success for theory, observation and humanity!

But the reader should also be {\it puzzled}: Most of the Universe (about 95\%!) is made of stuff that we have no idea about, namely (cold) dark matter and the cosmological constant (or {\it dark energy} in its more general albeit more alien form). {\it This cannot be right, or can it?}
And probably most importantly: {\it What made all this possible? How did we deduce all this? How can local measurements teach us about physical processes occurring at times and distances that we can barely fathom?}

The last question is highly profound and one of the central parts relating to CMB physics. Here, I will try to explain what information is encoded in the CMB frequency spectrum (Sect.~\ref{sec:spectral distortions}) and how it helped in building our detailed model of the Universe, while CMB anisotropy science is covered elsewhere. I will also give a broad brush overview of some of the unique future opportunities with CMB spectral distortions and how they could further help us gain even deeper insights into the fabric of the cosmos in the coming decades.

\section{The average CMB and spectral distortions}
\label{sec:spectral distortions}
Cosmology is about describing the average properties of the Universe at large scales. When thinking about the CMB, we can first study the average {\it frequency} spectrum, which characterizes how many CMB photons we receive at different observing frequencies, no matter which sky-direction we may look into. {\it What could we expect?} In full thermodynamic equilibrium, the photon distribution is given by that of a perfect blackbody radiator. We shall denote the blackbody {\it intensity} or {\it Planckian} as, $B_\nu(T)$, where $\nu$ is the observing frequency and $T$ the blackbody temperature. The Planckian is described by the Planck formula, which reads:
\beal{
\label{eq:planck_formula}
B_\nu(T)=\frac{2 h}{c^2}\frac{\nu^3}{\expf{h\nu/\kB T}-1}
}
with the common units $[B_\nu(T)]=\ergs  \sec^{-1}\cm^{-2} \Hz^{-1} \sr^{-1}= 10^{17}\,\MJysr$. Here, $h$ is Planck's constant, $c$ denotes the speed of light, $\kB$ is the Boltzmann constant, and $\nbb(\nu, T)=1/(\expf{h\nu/\kB T}-1)$ is the blackbody occupation number.

To reach equilibrium we require a process referred to as {\it thermalization}. Dense and hot environments, which imply very fast reactions between matter and radiation, will inevitably lead to efficient thermalization and hence a (local) blackbody spectrum. In the Big Bang model, these conditions are certainly met in the very early Universe, which, without any calculation, immediately leads to one important prediction: the average CMB spectrum should be close to that of a blackbody.
And indeed, since the measurements of \COBEF in the mid-90's we know that the CMB spectrum is extremely close\footnote{The CMB spectrum is often referred to as the most perfect blackbody in nature.} to that of a perfect blackbody at an average temperature $T_0\simeq 2.7255\,{\rm K}$, with departures from the Planckian -- the so-called {\it CMB spectral distortions} -- limited to one part in ten thousand.\footnote{Nobel Prize in Physics to John Mather in 2006 for this measurement.} This result not only provides strong support in favor of the Big Bang model, but also fixes the first key cosmological parameter -- $T_0$ -- to extremely high precision. In fact, the precision of this long-standing measurement some 30 years ago is so high that we often forget to even mention $T_0$ when talking about cosmology. 
However, without it we would have no accurate predictions from the BBN era or the cosmological recombination process, both crucial ingredients to our interpretation of the cosmological data, begging the question why we sometimes are so negligent!

After accepting the Big Bang picture as a given, {\it could we have expected anything else but a blackbody spectrum?} The answer is {\it yes}! First, while during the earliest phases of cosmic history high densities and temperatures are guaranteed to thermalize everything, what happens later when the Universe expands and cools down? Is the CMB spectrum really unaffected? Also, what if during the course of its evolution some {\it cosmic explosion} went off? Would the CMB spectrum really show no traces of this? Or put another way, {\it until when} can one perturb the equilibrium between matter and radiation without creating spectral distortion (i.e., non-equilibrium) signatures?

As I will show in the chapter, several early-universe processes {\it inevitably} lead to CMB spectral distortions (SDs) at a level that is within reach of present-day technology. This provides strong motivation for studying the physics of CMB spectral distortions and asking what these small signals might be able to tell us about the Universe we live in. 
Even today, the measurements from \COBEF provide some of the most stringent constraints on new physics models. Harvesting this fundamental science in several cases has only become possible recently with modern theoretical tools to predict SDs. Furthermore, we have entered a new phase on the experimental side that promises improved measurements of the CMB spectrum in the coming years, as I will try to highlight here.

\subsection{What is the thermalization problem all about?}
\label{sec:thermalization is what}
When considering the cosmological thermalization problem we are asking: {\it how was the present average CMB spectrum really created?} Assuming that everything starts off with a pure blackbody spectrum, the uniform adiabatic expansion of the Universe alone (absolutely no collisions and spatial perturbations implied here!) leaves the CMB spectrum unchanged -- a blackbody thus remains a blackbody at all times. 
However, any out-of-equilibrium processes leading to photon production/destruction or energy release/extraction should inevitably introduce momentary distortions to the CMB spectrum. 
This could be related to {\it reionization} and {\it structure formation}; {\it decaying} or {\it annihilating particles}; the {\it dissipation of primordial density fluctuations}; {\it cosmic strings, axions} and {\it dark radiation}; {\it primordial black holes}; {\it small-scale magnetic fields} and the {\it cosmological recombination process} to name some prominent examples. This list certainly makes theorists very happy, but more importantly, many of these processes (e.g., reionization and cosmological recombination) are part of our standard cosmological model and therefore should lead to {\it guaranteed signals} that we can predict and search for. This shows that studies of spectral distortions offer both the possibility to {\it constrain well-known physics} but also to open up a {\it discovery space} for non-standard physics, potentially adding new {\it time-dependent information} to the picture.

In all this, the big question is: {\it was there enough time from the creation of the distortion until today to fully restore the blackbody shape, pushing distortions below any observable level?} 
To restore equilibrium, we need to redistribute photons in energy. In the early Universe, this is mediated by the well-known Compton scattering process, in which a free electron scatters with a photon, $e+\gamma \rightarrow e' + \gamma'$, Comptonizing the spectrum. However, Compton scattering conserves the number of photons and alone is {\it unable} to fully restore a blackbody spectrum. To see this, imagine that some process injected energy into the photon field, raising the average photon energy density to $\rho_\gamma' = \rho_\gamma+\Delta \rho_\gamma$. After some time, the spectrum, which started as a blackbody at a temperature $T$, will have relaxed to a new stationary distribution under repeated Compton scattering, the aforementioned Comptonization process. By assuming that this new distribution is again a blackbody, we can easily compute the new blackbody temperature as $T'_\rho=T(1+\Delta \rho_\gamma/\rho_\gamma)^{1/4}$, knowing that $\rho_\gamma \propto T^4$ for blackbody radiation.\footnote{Here we assumed that the other particles that couple to the photons do not take any significant share of the injected energy once a stationary solution is reached. In our Universe, this is indeed fully justified after the BBN and electron-positron annihilation era, i.e., temperatures $\kB T\lesssim 10\,\keV-50\,\keV$ or at redshift $z \lesssim 10^7$.} Here, the subscript '$\rho$' emphasizes that the effective photon temperature is deduced from the energy density of the photon field. Similarly, by counting photons we can compute the number density based temperature, $T_N=T(1+\Delta N_\gamma/N_\gamma)^{1/3}$, realizing that for a blackbody the number density follows $N_\gamma \propto T^3$.
However, in our example, by construction we have increased the photon energy density but not changed the photon number density, i.e., $\Delta N_\gamma=0$. We thus conclude that $T_N=T<T_\rho$, which implies that the new stationary photon distribution {\it cannot} be a blackbody, no matter how long we continue scattering photons.

{\it What is missing?} We also need to adjust the photon number, which in the cosmos happens mainly through double Compton emission with corrections from Bremsstrahlung. 
Bremsstrahlung (or the free-free process) is well-known from standard textbooks on astrophysical processes. But what is double Compton scattering? Photons Compton scattering doubly? -- Far off. The double Compton process is the first radiative correction to the Compton process, where a soft photon, $\gamma''$, with photon frequency $\nu''\ll \nu$, is created in the collision: $e+\gamma \rightarrow e' + \gamma'+\gamma''$. This process allows adjusting the number of photons and can be computed using quantum electrodynamics. In fact, Bremsstrahlung is nothing but the first radiative correction to the Coulomb scattering process between electrons and ions, so we are quite familiar with these kind of problems and know how to calculate them.

{\it But why does double Compton dominate when we rarely hear about it?} This boils down to an extremely important property of the Universe: its immense excess of photons over baryons, with a photon number density, $N_\gamma$, that exceeds the baryon number density, $\Nbary$, more than a billion times, $N_\gamma \simeq \pot{1.63}{9}\,\Nbary$. This fact also causes BBN and the cosmological recombination process to end at much lower temperature than is naively expected, simply because there are so many energetic photons per baryon in the Wien tail for the CMB. The production rate of photons by Bremsstrahlung scales as $\propto \Ne N_{\rm i} f_{\rm BR}(T, \nu)$ with the number density of electrons and ions, $\Ne$ and $N_{\rm i}$, while for double Compton one expects $\propto \Ne N_{\gamma} f_{\rm DC}(T, \nu)$. Here, $f_{\rm BR}(T, \nu)$ and $f_{\rm DC}(T, \nu)$ determine the precise temperature and frequency dependence of the respective processes, but even without knowing their precise form, we have at least nine orders of magnitudes to play with from the ratios of number densities of colliding particles, making double Compton emission an important process to consider.

By understanding the thermalization problem and studying the CMB spectrum in fine detail we can thus learn about different early-universe processes and the thermal history of our Universe.
This can open a new window to the early Universe, allowing us to peek behind the {\it last scattering surface}, which is so important for the formation of the CMB temperature and polarization anisotropies \citep[e.g., see][within this series]{Rahimi2024EMMS}. SDs are thus a great probe of physics in the primordial Universe.

\subsection{General conditions relevant to the thermalization problem}
\label{sec:conditions}
In the early Universe, photons undergo many interactions with the other particles. We shall mainly concern ourselves with the {\it average} CMB spectrum and neglect temperature anisotropies and density fluctuations when describing distortion evolution. Distortion anisotropies can be created through anisotropic energy release processes.\footnote{Some obvious distortion anisotropies are created by galaxy clusters through the Sunyaev-Zeldovich at late times. Similar physics can occur in the early Universe as we explain briefly towards the end of this chapter.} However, these are usually very small, such that we only briefly touch on them later. For now, we also assume that the distortions are always minor in amplitude, so that the problem can be linearized. This allows us to resort to a Green's function approach when solving the thermalization problem, which greatly simplifies explicit thermalization calculations for different energy release scenarios as can be carried out using full thermalization codes such as\footnote{A list of some of the relevant codes is given at the end of this chapter.} {\tt CosmoTherm}.

We furthermore assume the standard $\Lambda$CDM background cosmology with standard ionization history computed using the cosmological recombination code {\tt CosmoRec}. The electron and baryon distribution functions are modeled by Maxwellians at a common temperature, $\Te$, down to very low redshifts ($z\lesssim 10$), when the thermalization process is already extremely inefficient. We furthermore need not worry about the evolution of distortions before the electron-positron annihilation era ($z\gtrsim 10^7-10^8$), since in this regime rapid thermalization processes always ensure that the CMB spectrum is extremely close to that of a blackbody. 
With these conditions in mind, we are therefore just dealing with non-relativistic electrons ($\kB\Te\lesssim 1\,\keV$), protons and helium nuclei immersed in a bath of CMB photons. We can also neglect the traces of other light elements for the thermalization problem and usually assume that neutrinos and dark matter are only important for determining the expansion rate of the Universe.
This implies that the cosmological thermalization problem boils down to a purely time-dependent problem describing the evolution of the CMB spectrum at every frequency.

\section{Photon Boltzmann equation for the average spectrum}
\label{Ch:Boltzmann_eq}
All investigations of the formation and evolution of CMB, both spatially and in frequency, begin with the radiative transport or {\it Boltzmann equation} for the photon phase space distribution, $n(x^\mu, p^\mu)$. Often $n(x^\mu, p^\mu)$ is also referred to as the photon occupation number, where $x^\mu$ and $p^\mu$ denote the position and momentum four-vectors. It is related to the intensity of the photon field, $I$, with $p=h\nu/c$, as $$n=\frac{c^2 I}{2h\nu^3}.$$ 
This means that for a blackbody one has $\nbb(\nu, T)=c^2 B_\nu/[2h\nu^3]=(\expf{h\nu/\kB T}-1)^{-1}$. The occupation number is a very convenient choice for the distribution function, due to useful properties such as its Lorentz invariance. It also highlights an important aspect of Bosons, which in contrast to Fermions tend to bunch up in regions of high occupation as can be found at $h\nu \ll \kB T$, where one has $\nbb(\nu, T)\simeq \kB T/[h\nu]$. In reality, the physics of the $\mu$-distortion depends on this property (see Sect.~\ref{sec:mu_dist_section}).

Here, we are only interested in the evolution of the average spectrum assuming no spatial perturbations. In this case, spatial fluctuations can be neglected, such that $n(x^\mu, p^\mu)\rightarrow n(t, p)$ and we may express the average photon Boltzmann equation as
\beal{
\label{eq:gen_Boltzmann_equation_II_final}
\frac{\partial n}{\partial t}- H \,p \frac{\partial n}{\partial p}={\rm C}[n],
}
suppressing the explicit arguements. Here, $H(t)=\id\ln a/\id t= \dot a/a$ is the standard Hubble expansion rate given the scale factor, $a$, and ${\rm C}[n]$ denotes the so-called Boltzmann collision term, which accounts for interactions of photons with the other species in the Universe. The collision term incorporates several important effects. Most importantly, Compton scattering couples photons and electrons, keeping the two in close thermal contact until low redshifts, $z\lesssim 100-200$, without changing the number of photons. Bremsstrahlung and double Compton emission allow adjusting the photon number and are especially rapid at low frequencies, establishing a Rayleigh-Jeans spectrum at the temperature of the electrons, as we explain below. We can also have extra photon source terms that depend on other processes.

\vspace{-2mm}
\subsection{Solution without collisions}
\label{Ch:Boltzmann_eq_no_colls}
Without specifying the form of the collision term, not much progress can be made. Let us first consider the case when {\it no collisions} are present (${\rm C}[n]\equiv 0$). By transforming to comoving photon momentum, $\tilde{p}=a(t) p$, we have $p=\tilde{p}/a$ and $$\partial_t n(t, p)\big|_p=\partial_t n(t, \tilde{p}/a)\big|_p=\partial_t n(t, \tilde{p})\big|_{\tilde{p}}+\partial_{p} n(t, p)\big|_t\,\tilde{p} \,\partial_t (-1/a)=\partial_t n(t, \tilde{p})\big|_{\tilde{p}}+p \,\partial_{p} n(t, p)\big|_t  \,H.$$ Using this conversion, from Eq.~\eqref{eq:gen_Boltzmann_equation_II_final} we then have $\partial_t n(t, \tilde{p})\big|_{\tilde{p}}\equiv 0$, implying that the photon occupation number is conserved and given by the initial distribution through, $n(t, p)=n[t_{\rm i}, p \,a(t)/a(t_{\rm i})]$. In the absence of collisions only the photon momenta are redshifted but the shape of the spectrum is unaltered. A blackbody remains a blackbody as we already anticipated! 

We will see that even in the simplest case, the collision term does {\it not} vanish because baryons cool faster than photons and thereby continuously extract some amount of energy from the photon bath, creating a small but inevitable spectral distortion (see Sect.~\ref{sec:ad_cool}). However, it is still extremely useful to absorb the Hubble expansion term by going to comoving coordinates for thermalization calculations.\footnote{Even for numerical treatments of the thermalization problem this substitution is extremely useful since otherwise certain boundary conditions can cause problems.} 
It is customary to introduce the dimensionless frequency variable $x=h\nu/\kB T_z(t)$, with $\nu=c p/h$ and  $T_z(t)=T_z(t_{\rm i})\,a(t_{\rm i})/a(t)\propto (1+z)$. The average photon Boltzmann equation then takes the slightly more compact form 
\beal{
\label{eq:gen_Boltzmann_equation_II_final_comov}
\frac{\partial n(t, x)}{\partial t}={\rm C}[n(t, x)],
}
which explicitly highlights the conservation law in the absence of collisions. We highlight here that the value of the reference temperature, $T_z(t)$, given above is not at all crucial for obtaining the form Eq.~\eqref{eq:gen_Boltzmann_equation_II_final_comov}. All that is required is that $T_z(t) \propto a^{-1}$. As such, $T_z(t)$ should not be confused with the effective photon temperature, $\Tg$, or the CMB temperature, $\TCMB=T_0(1+z)$, although it is often useful to choose it based on the properties of the photon field.

\vspace{-2mm}
\subsection{Collision term for Compton scattering}
\label{sec:CS_phys}
We already mentioned that Compton scattering is responsible for redistributing photons in energy. This problem has been studied a lot in connection with X-rays from compact objects but also in the cosmological context. In reality, electron-photon scattering also helps isotropizing (i.e., spatially thermalizing) the photon field (in the Thomson scattering limit), although for this energy exchange is not crucial. 

To account for the Comptonization of photons by free thermal electrons, we can use the so-called {\it Kompaneets equation}:\footnote{
The Kompaneets equation can be obtained by computing the Compton collision term in the limit $h\nu\ll k\Te$ and $k\Te \ll \me c^2$, keeping only terms up to first order in $\The$ and $h\nu/\me c^2$ (a so-called Fokker-Planck expansion). This is equivalent to considering the first two moments of the photons frequency shift, $\Delta \nu/\nu$, over the scattering kernel. 
}
\beal{
\label{eq:Kompaneets}
\left.\frac{\partial n}{\partial \tau}\right|_{\rm CS}
&\approx \frac{\The}{\xe^2}\frac{\partial}{\partial \xe} \xe^4 \left[\frac{\partial}{\partial \xe} n +n(1+n) \right]
\equiv \frac{\The}{x^2}\frac{\partial}{\partial x} x^4 \left[\frac{\partial}{\partial x} n +\frac{\Tz}{\Te}\,n(1+n) \right],
}
where $\id \tau = \Ne \sigT c \id t$ is the Thomson optical depth with Thomson scattering cross section, $\sigT$; $\The=k\Te/\me c^2$ is the dimensionless electron temperature in units of the electron rest mass energy, $\me c^2$, and $\xe=h\nu/k\Te\neq x$ is an alternative frequency variable, where the electron temperature, $\Te$, is used as a reference.
The Kompaneets equation can be used to describe the repeated scattering of photons by thermal electrons in isotropic media. The first term in the brackets describes {\it Doppler broadening} and {\it Doppler boosting} and the second term accounts for the {\it recoil effect} ($\propto n$) and {\it stimulated recoil} ($\propto n^2$) . The latter terms are especially important for reaching full equilibrium in the limit of many scatterings, as we will see below.

In the next sections, we discuss analytic solutions of the Kompaneets equation in limiting cases. Here, a couple of words about limitations of this equation are in order. First of all, we assumed that the change in the energy of the photon in each scattering event is small. For hot electrons ($\kB \Te \gtrsim 5\,\keV$) this is no longer correct and one has to go beyond the lowest orders in $\Delta \nu/\nu$. This is for example important for the {\it Sunyaev-Zeldovich effect} \citep{Zeldovich1969} of very hot clusters, causing so-called relativistic temperature corrections \citep{Sazonov1998, Challinor1998, Itoh98, Chluba2012SZpack}. The second limitation is that if the photon distribution has sharp features (more narrow than the width of the scattering kernel) then the shape of the scattered photon distribution is not well represented with the diffusion approximation. In this case, a scattering kernel approach can be used to describe the scattering problem, although developing efficient numerical schemes to describe many scatterings can become cumbersome.
However, for the conditions in the early Universe after the BBN era, we are usually safe and can apply the Kompaneets equation.

\begin{figure}
\centering
\includegraphics[width=0.72\columnwidth]{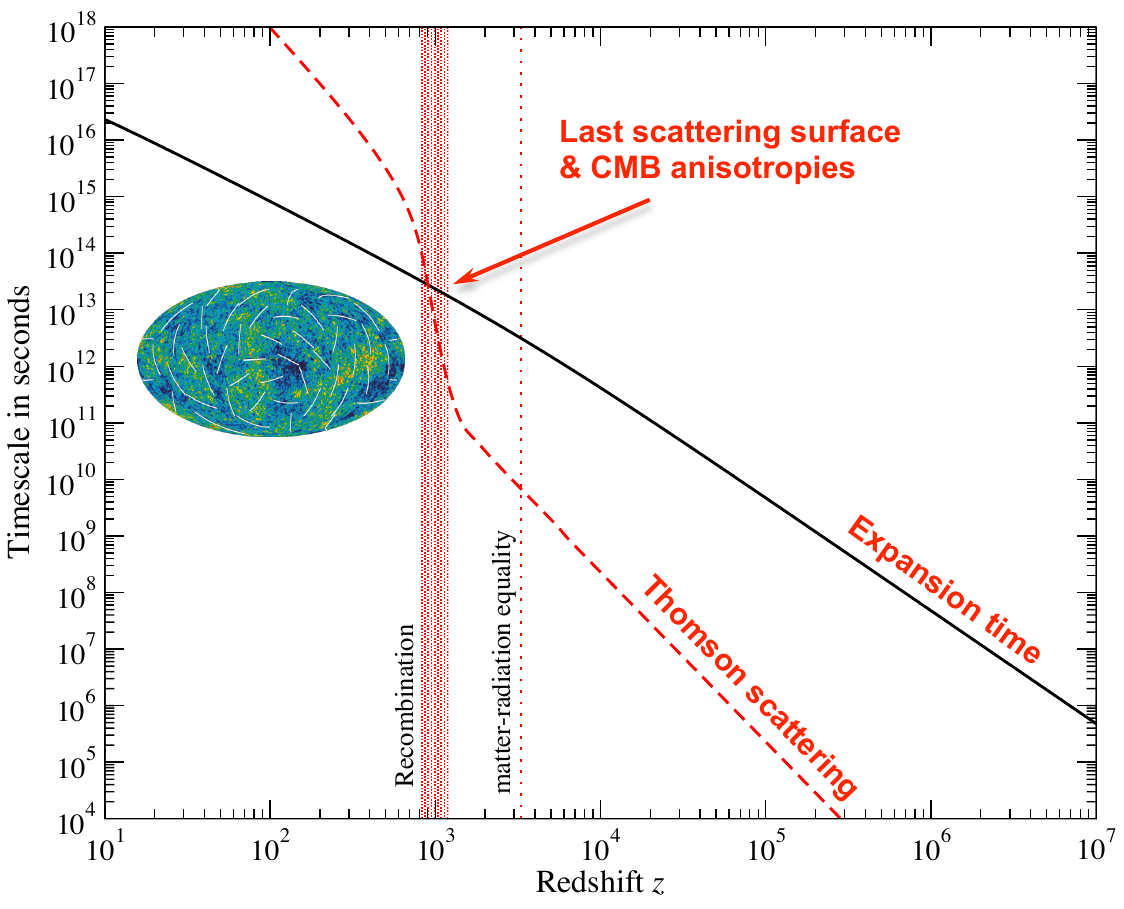}
\caption{Comparison of the Thomson scattering timescale with the Hubble expansion timescale. The Thomson scattering timescale exceeds the expansion timescale during the recombination era, when the Universe becomes transparent. This is the era when the CMB anisotropies become visible. Figure is taken from \citet{Chluba2018Varenna}.}
\label{fig:Thomson_Hubble_times}
\end{figure}
\vspace{-2mm}
\subsubsection{Comptonization efficiency}
From the Kompaneets equation, we can already understand some of the important aspects of Comptonization, by simply looking at characteristic timescales. One important quantity is the Thomson scattering timescale, $t_{\rm T}=(\sigT \Ne c)^{-1}$. It describes how rapidly photons scatter with free electrons. For the standard cosmology with $24\%$ of helium (by mass), we have
\vspace{-1mm}
\beal{
\label{eq:t_Thomson}
t_{\rm T}=(\sigT \Ne c)^{-1} \simeq \pot{2.7}{20} \,X_{\rm e}^{-1} (1+z)^{-3} \, \sec \simeq  \pot{4.0}{4} \left[\frac{X_{\rm e}}{0.16}\right]^{-1}\left[\frac{1+z}{1100}\right]^{-3}\,{\rm years},
}
where $X_{\rm e}=\Ne/N_{\rm H}$ is the free electron fraction relative to the total number of hydrogen nuclei, $N_{\rm H}$. At $z=1100$ (where $X_{\rm e}\approx 0.16$), this corresponds to $\simeq 40\,000$ years between scatterings! What does this really mean? 

To put this number into perspective we have to compare it with the typical expansion timescale given by the inverse Hubble rate:
\vspace{-1mm}
\beal{
\label{eq:t_exp}
t_{\rm exp}=H^{-1} \simeq 
\begin{cases}
\pot{4.8}{19} \,(1+z)^{-2} \, \sec &\qquad \text{(radiation domination)}
\\
\pot{8.4}{17} \,(1+z)^{-3/2} \, \sec  &\qquad \text{(matter domination)},
\end{cases}
}
where the transition between matter and radiation (photons + neutrinos) domination occurs around $\zeq\simeq 3400$. From Fig.~\ref{fig:Thomson_Hubble_times} we see that the Thomson scattering rate (shorter timescales) is much higher than the Hubble expansion rate until after decoupling around $z\simeq 10^3$. But even then, the timescale for scattering only exceeds the expansion time by a factor of $\simeq 10^2-10^4$ at $z\lesssim 10^3$. However, this is when the isotropization process of CMB anisotropies becomes inefficient and we start seeing the primordial CMB fluctuations.

The most important aspect of Comptonization is energy exchange between electrons and photons. The timescale on which electrons transfer energy to the photons is
\vspace{-1mm}
\beal{
\label{eq:t_egamma}
t_{\rm e \gamma} \approx \frac{t_{\rm T}}{4 \The} \simeq \pot{4.9}{5} \, t_{\rm T} \,\left[\frac{1+z}{1100}\right]^{-1}\simeq \pot{1.2}{29} (1+z)^{-4}\,\sec.
}
How can one see this? In simple words, the timescale for scattering is $t_{\rm T}\simeq [\Ne \sigT c]^{-1}$ and per scattering the fractional energy-exchange between photons and electrons is $\Delta \nu/\nu\simeq 4 \The$. The photon energy therefore changes as $\dot{E}/E\approx 4 \The \, t^{-1}_{\rm T}$, yielding the desired estimate.

\begin{figure}
\centering
\includegraphics[width=0.68\columnwidth]{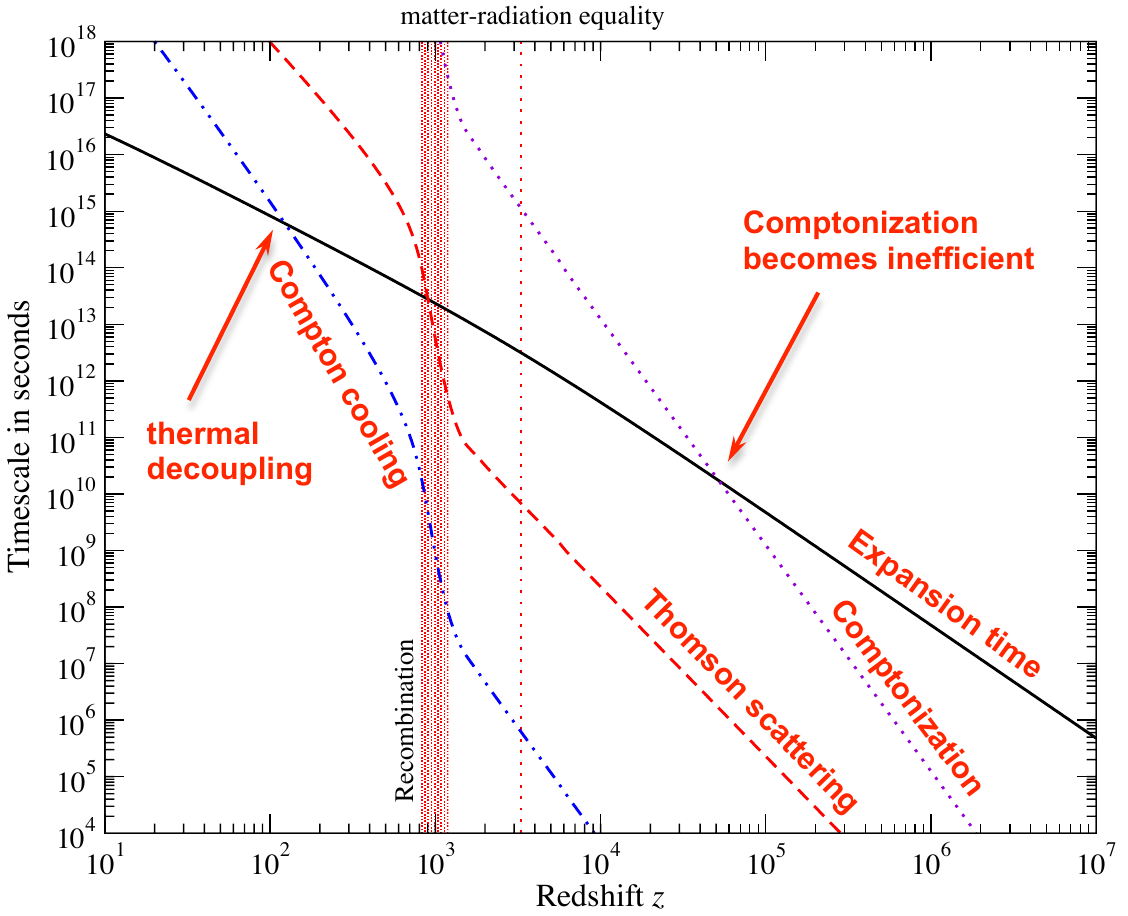}
\caption{Comparison of the Comptonization,  Compton cooling, Thomson scattering and Hubble expansion timescales. The character of the distortion is expected to change around $z\simeq 50,000$, when Comptonization becomes inefficient. Figure is taken from \citet{Chluba2018Varenna}.}
\label{fig:Compton_Hubble_times}
\end{figure}

By comparing $t_{\rm e\gamma}$ with the Hubble rate one finds that at redshift $z_{\mu y}\simeq \pot{5}{4}$, {\it Comptonization} becomes inefficient (see Fig.~\ref{fig:Compton_Hubble_times}). This is a very important redshift, since the characteristic spectral distortions is expected to change, transitioning from a so-called $\mu$-distortion (with extremely efficient energy exchange between photons and electron) to a $y$-type distortion (with inefficient energy exchange). Evidently, the transition is not abrupt and the characteristic shape of the distortion adjusts over a range of redshift between $z\simeq 10^4 - \pot{3}{5}$, but we shall explain all these details in the next few sections.

The Comptonization timescale is quite long compared to the timescale over which electrons are heated by photons. The big difference is that every electron has $\simeq \pot{1.9}{9}$ photons to scatter with, making the number of interactions much larger. As already mentioned before, this fact influences many phases in the history of the Universe. For example, the cosmological recombination process is delayed until the temperature of the CMB has dropped below $k\Tg \simeq 0.26\,{\rm eV}$, which is $\simeq 50$ times smaller than the ionization potential of hydrogen, $E_{\rm ion}\simeq 13.4\,{\rm eV}$. Similarly, BBN occurs significantly later than is expected from naively considering nuclear binding energies.

By comparing the non-relativistic thermal energy of the plasma, $\rho_{\rm th} = (3/2) \sum_i N_i k \Te=(3/2) N_{\rm H} (1+f_{\rm He}+X_{\rm e})\,k \Te$, with the energy density of photons, $\rho_\gamma\approx 0.26\,\eV \cm^{-3}\,(1+z)^4$, we have 
\vspace{-1.5mm}
\beal{
\label{eq:t_gammae}
t_{\rm \gamma e} = \frac{\rho_{\rm th}}{\rho_\gamma} \, t_{\rm e \gamma} 
=\frac{3 N_{\rm H} (1+f_{\rm He}+X_{\rm e})}{8\rho_\gamma/(\me c^2)} \, t_{\rm T} 
\simeq 0.31 \, t_{\rm T} \,(1+z)^{-1}\simeq \pot{7.3}{19} (1+z)^{-4}\,\sec.
}
Here, we used $X_{\rm e}=1+2f_{\rm He}\approx 1.16$ (fully ionized hydrogen and helium plasma), $f_{\rm He}\approx \Yp/[4(1-\Yp)]\approx 0.079$ ($\Yp=0.24$) and the number density of hydrogen nuclei, $N_{\rm H}=(1-\Yp)\,\Nbary$.  
Before recombination, the Compton cooling time is about $\simeq \pot{1.6}{9}$ times shorter than the Comptonization time. This means that electrons and baryons (through Coulomb scatterings) remain in full thermal contact with the photon field until very late. From Fig.~\ref{fig:Compton_Hubble_times} one can see that thermal decoupling is expected to happen somewhere around $z\simeq 100-200$.

\vspace{-1.5mm}
\subsection{Bremsstrahlung and double Compton emission}
So far we have only considered the redistribution of photons in energy by the Compton process. As explained above, this alone is insufficient for thermalizing the radiation field. In addition, we need to adjust the photon number, which in the expanding early Universe is achieved by thermal Bremsstrahlung (BR) and double Compton (DC) emission. In our Universe, DC emission is much more important for the cosmological thermalization process than BR. Nevertheless, at late times BR has to be included for accurate computations. 

The collision term for BR and DC emission is quite complicated but can be derived using standard quantum electrodynamics and astrophysics textbooks \citep{Rybicki1979, Thorne1981, Lightman1981}. It can be 
compactly expressed as
\bsub
\label{eq:dn_em_abs}
\beal{
\left.\frac{\partial n(\tau, x)}{\partial \tau}\right|_{\rm em/abs}
&= 
\frac{K_{\rm BR}\,\expf{-\xe}+K_{\rm DC}\,\expf{-2x}}{x^3}\!\left[ 1 - n(\tau, x) \, (\expf{\xe}-1)\right]
\\[0.5mm]
K_{\rm BR}
=\frac{\alpha}{2\pi}\frac{\lambda_{\rm e}^3}{\sqrt{6 \pi}\,\The^{7/2}}\,\left(\frac{\Te}{\Tz}\right)^3 \sum_i Z_i^2 &N_i \, \bar{g}_{\rm ff}(Z_i, \Te, \Tz, \xe),
\quad
K_{\rm DC}=\frac{4\alpha}{3\pi}\,\Thz^2\,I_{\rm dc} \,g_{\rm dc}(\Te, \Tz, x)
\\[0.5mm]
&\!\!\!\!\!\!\!\!\!\!\!\!\!\!\!\!\!\!\!\!\!\!\!\!\!\!\!\!\!\!\!\!\!\!\!\!\!\!\!\!\!\!\!\!\!\!\!\!\!\!\!\!\!\!\!\!\!\!\!\!\!\!\!\!\!\!
\bar{g}_{\rm ff}(\xe)\approx 
\begin{cases}
\frac{\sqrt{3}}{\pi}\ln\left(\frac{2.25}{\xe}\right)&\text{for}\quad \xe\leq 0.37
\\
1 &\text{otherwise}
\end{cases},
\qquad
g_{\rm dc}\approx 
\frac{1+\frac{3}{2}x+\frac{29}{24} x^2+\frac{11}{16} x^3+\frac{5}{12} x^4}{1+19.739\Thz-5.5797\Thz}.
}
\esub
where $\alpha\approx 1/137$ is the fine structure constant, $I_{\rm dc}=\int x^4 n (1+n)\id x\approx 4\pi^4/15$ is the DC Compton emissivity integral of a blackbody at a temperature $\Tz$ (or $\Thz=\kB\Tz/\me c^2$), and $\lambda_{\rm e} =h/\me c\simeq \pot{2.43}{-10}\,\cm$ is the Compton wavelength of the electron. We furthermore denote the BR and DC Gaunt factors as $\bar{g}_{\rm ff}$ and $g_{\rm dc}$, respectively, with some simple approximations that are useful for estimates. These capture quantum-mechanical corrections to the classical treatments of these process and can be efficiently computed using {\tt BRpack} and {\tt DCpack}. 

From Eq.~\eqref{eq:dn_em_abs} one can already see that, as expected, both BR and DC push the radiation field into equilibrium with a blackbody at the temperature of the electrons, $n_{\rm e}=1/(\expf{\xe}-1)$. Due to the $1/x^3$ scaling of the emissivity it is clear that BR and DC emission are both most important at low frequencies, where a blackbody spectrum is extremely quickly restored. Inserting typical numbers for $z\gtrsim 10^3$ and assuming $\Te\approx \Tz \simeq \TCMB$, we have 
\vspace{-1.5mm}
\bsub
\beal{
\label{eq:BR_DC_estimates}
K_{\rm BR}&\simeq \pot{1.4}{-6}\,\left[\frac{\bar{g}_{\rm ff}}{3.0}\right] \left[\frac{\Omega_{\rm b} h^2}{0.022}\right]\,(1+z)^{-1/2}
\\[0mm]
K_{\rm DC}&\simeq \pot{1.7}{-20}\,(1+z)^{2}.
}
\esub
This implies that at $z_{\rm dc,br}\simeq \pot{3.7}{5}\left(\left[\frac{\bar{g}_{\rm ff}}{3.0}\right] \left[\frac{\Omega_{\rm b} h^2}{0.022}\right]\right)^{2/5}$ BR and DC emission are similarly important. At $z>z_{\rm dc,br}$, DC emission is much more crucial, while at lower redshifts BR dominates. The DC process is thus extremely important for forming the CMB blackbody spectrum.

\subsection{Final evolution equations}
We now have all the pieces to write down the required photon evolution equation including Compton scattering, double Compton scattering and Bremsstrahlung, and also an external source term: 
\beal{
\label{eq:photon_Boltz}
\frac{\partial n(\tau, x)}{\partial \tau}
&= 
{\rm C}[n]_{\rm CS}+{\rm C}[n]_{\rm em/abs}+{\rm C}[n]_{\rm S}
\equiv
\frac{\The}{x^2}\frac{\partial}{\partial x} x^4 \left[\frac{\partial}{\partial x} n +\frac{\Tz}{\Te}\,n(1+n) \right]
+
\frac{K\,\expf{-\xe}}{x^3}\!\left[ 1 - n \, (\expf{\xe}-1)\right]
+S(\tau, x),
}
where $K=K(\tau, x, \xe, \Te, \Tz)=K_{\rm BR}\,+K_{\rm DC}\,\expf{\xe-2x}$ determines  effective DC+BR photon production rate. This equation has to be supplemented by the evolution equation for the electron's temperature. The evolution of $\Te$ is especially important since energy release often only affects the electrons / baryons which then become hotter to up-scatter CMB photons via Compton scattering. Here, we shall neglect heating or cooling through DC and BR. These contributions are not as important for the evolution of distortions from external heating processes although they can lead to extremely interesting effects when {\it soft photon sources} are being studied in the post-recombination era ($z\lesssim 10^3$). 

The evolution equation for the electron (and baryon) temperature then takes the simple form
\beal{
\label{eq:temperature}
\alpha_{\rm h}\frac{\id \Te}{\id \tau}
&\approx - 2 H \Te \, \alpha_{\rm h} \, t_{\rm T}+\frac{4 \rho_\gamma}{\me c^2}\left[\Te^{\rm eq}-\Te\right]+\frac{\id  Q}{\id \tau}
}
where $\alpha_{\rm h}=\frac{3}{2}\kB N_{\rm H}\,\left[1+f_{\rm He}+X_{\rm e}\right]$ is the heat capacity of the baryons and $\id  Q/\id \tau$ is the comoving heating term (obtained from the relevant collision term), which could be due to decaying particles, magnetic fields or other sources of heating, as we discuss below. We furthermore defined the {\it Compton equilibrium temperature}
\beal{
\label{eq:Compton_eq_temperature}
\Te^{\rm eq}&=\Tz \frac{\int x^4 n(1+n) \id x}{4\int x^3 n \id x},
}
which usually is $\Te^{\rm eq}\neq \Tz$ for a distorted radiation field. The first term in Eq.~\eqref{eq:temperature} captures the effect of the Hubble expansion on the matter temperature, while the second term is due to the Compton scattering of electrons with photons. 

Solving the coupled system of equations, Eq.~\eqref{eq:photon_Boltz} and \eqref{eq:temperature}, generally is only possible numerically. However, today we have thermalization codes such as {\tt CosmoTherm} to obtain the solution for a wide range of scenarios, even going beyond some of the approximations that we made above. The general strategy is to discretize the photon distribution into frequency bins (usually a couple of thousand) and then convert the problem into a system of coupled ordinary differential equations. With modern computers, one can obtain solutions for one scenario in about a minute. Since the present-day distortions are known to be small, one can often linearize the problem. This allows us to compute the Green's function, which then paves the path towards fast representations of the solutions for various heating scenarios. 
We will now consider some solutions of the thermalization problem in limiting cases. This will reveal the main distortion signals and also provides an understanding of important aspects of distortion physics, answering some of the questions posed at the beginning of the section.

\section{Types of spectral distortions from energy release}
We are now in the position to discuss the main types of spectral distortions created by energy release. In Sect.~\ref{sec:CS_phys}, we learned that around $z_{\mu y}\simeq \pot{5}{4}$ the Comptonization timescale ($\leftrightarrow$ transfer of energy from electrons to photons) becomes longer than the Hubble time. It is clear that this marks an important transition in the efficiency of Compton scattering and the redistribution of photons in energy. Let us try to quantify this a little better by looking at the photon evolution equation, for now neglecting photon emission
\beal{
\label{eq:CS_Kompa_limit}
\frac{\partial n}{\partial \tau}
\approx \frac{\The}{x^2}\frac{\partial}{\partial x} x^4 \left[\frac{\partial}{\partial x} n +\frac{\Tz}{\Te}\,n(1+n) \right],
}
and setting $n=n(\tau, x)$. This equation has no general analytic solution, but we can solve it for limiting cases. As we explain next, a Compton-$y$ distortion is created by scatterings with inefficient energy exchange between electrons and photons, while a chemical potential $\mu$-distortion is formed in the regime of extremely efficient energy exchange. At intermediate regimes, time-dependent corrections to these extremes are introduced, in principle allowing us to extract detailed information about the physical process that created the distortions by measuring the precise distortion shape (see Sect.~\ref{sec:Transition}). Overall, this makes CMB SDs a powerful probe of the thermal history of the Universe, with the potential to constrain both standard and new physics scenarios.

\subsection{Scattering of CMB photons in the limit of small $y$-parameter}
Assuming that at $\tau=0$ we start with a blackbody spectrum $n=n_{\rm bb}=1/(\expf{x}-1)$ at a temperature $\Tz$, we can estimate the distortion that is created after a very short time $\Delta \tau\ll 1$ as\footnote{In numerical terms, this is a simple explicit Euler step.} $\Delta n \approx \Delta \tau \,{\rm C}[n_{\rm bb}]$. Inserting $n_{\rm bb}$ into Eq.~\eqref{eq:CS_Kompa_limit}, we then find
\beal{
\Delta n
&\approx   \frac{\Delta \tau \,\The}{x^2}\frac{\partial}{\partial x} x^4 \left[\frac{\partial}{\partial x} n_{\rm bb} +\frac{\Tz}{\Te}\,n_{\rm bb}(1+n_{\rm bb}) \right]
= 
\frac{\Delta \tau\,(\Thz - \The)}{x^2}\frac{\partial}{\partial x} x^4 n_{\rm bb}(1+n_{\rm bb})
\nonumber \\[1mm]
&=  
\Delta \tau \,(\Thz - \The) \left[ 4 x n_{\rm bb}(1+n_{\rm bb}) - x^2 n_{\rm bb}(1+n_{\rm bb}) (1+2 n_{\rm bb})\right]
\nonumber \\[1mm]
&=  
\Delta \tau\,(\The - \Thz) \,\frac{x\expf{x}}{(\expf{x}-1)^2}\,\left[ x \,\frac{\expf{x}+1}{\expf{x}-1}-4\right]\equiv  \Delta \tau\, (\The - \Thz)\,Y(x),
}
where we used $\partial_x n_{\rm bb}=-n_{\rm bb}(1+n_{\rm bb})=-\expf{x}/(\expf{x}-1)^2$, $(1+2 n_{\rm bb})=(\expf{x}+1)/(\expf{x}-1)=\coth(x/2)$ and $\Thz=\kB\Tz/\me c^2$. The last line is the definition of the so called {\it Compton-$y$} distortion 
\beal{
\label{eq:def_YSZ}
Y(x)&=\frac{x\expf{x}}{(\expf{x}-1)^2}\,\left[ x \,\frac{\expf{x}+1}{\expf{x}-1}-4\right],
}
which arises in the limit of scatterings with inefficient energy exchange. The $y$-distortion of the CMB was first studied by \citet{Zeldovich1969} and then applied to hot electrons residing inside the potential wells of clusters of galaxies, giving rise to the so-called {\it thermal Sunyaev-Zeldovich (SZ) effect}. The important variable is the {\it Compton-$y$ parameter}
\beal{
\label{eq:def_y_parameter}
y=\int_0^\tau \frac{k(\Te-\Tz)}{\me c^2} \id \tau'=\int_0^t \frac{k(\Te-\Tz)}{\me c^2} \sigT \Ne c \id t',
}
which depends on the number of scatterings (related to $\tau$) and the net energy exchange\footnote{To some extent it would be better to immediately write $y^*=\int_0^\tau 4\frac{k(\Te-\Tz)}{\me c^2} \id \tau'$, so that $y^*=4y=\Delta \rho_\gamma/\rho_\gamma$ evidently gives the total amount of energy transfer.}, $\Delta \nu/\nu\simeq 4(\The-\Thz)\ll 1$, per scattering. Clearly, for $\Te\equiv \Tz$ one has $y=0$ and $\Delta n=0$, no matter how many scatterings actually take place! The solution Eq.~\eqref{eq:def_YSZ} for the distortion is thus valid as long as one has $|y|\ll 1$. This also ensures that the electron temperature does not change much by the scattering. 

One possible way to violate this condition even if the number of scatterings is tiny ($\tau\ll 1$) is by having a very large difference in the electron and photon temperature. Note that we also require $\The\ll 1$  since otherwise relativistic corrections to the Compton process appear, which are not accounted for by the Kompaneets equation. For the cosmological thermalization problem, we are always in the situation that the scattering $y$-parameter, $y_{\rm sc} = \int \Thz \id \tau$, is increased beyond unity by increasing the number of scatterings. In this case, Compton scattering pushes electrons and photons into kinetic equilibrium until a $\mu$-distortion is formed (see Sect.~\ref{sec:mu_dist_section}).

Assuming that we are in the regime $|y|\ll 1$, there are two cases of interest:
\begin{itemize}
\setlength{\itemindent}{\listindentation}
\vspace{2mm}
\item $y>0$: overall energy is transferred from the electrons to the photons \hspace{5mm} $\rightarrow$ \hspace{5mm}{\it Comptonization / photon up-scattering}

\item $y<0$: energy flows from the photons to the electrons \hspace{23.2mm} $\rightarrow$ \hspace{5mm}{\it Compton cooling / photon down-scattering}
\vspace{2mm}
\end{itemize}
In our Universe, $y>0$ is usually relevant, since most processes tend to heat the baryons. Therefore {\it negative $y$-distortions} are usually not being considered; however, the adiabatic cooling of matter in the expanding Universe (in the absence of extra heating) allows $\Te\lesssim \Tg$, so that $y<0$ does occur. This case is discussed in Sect.~\ref{sec:ad_cool}.

\subsection{Scattering of CMB photons in the limit of large scattering $y$-parameter}
\label{sec:mu_dist_section}
We now understand that for inefficient energy exchange between electrons and photons (i.e., $y\ll 1$) the shape of the distortion is determined by the $y$-parameter and has a spectral dependence, $Y(x)=G(x)[x \coth(x/2)-4]$. Let us next consider the other extreme, when many scatterings take place and the redistribution of photons in frequency is very efficient (i.e., $y\gg 1$). In the early Universe, this regime is found at redshifts $z\gtrsim \pot{5}{4}$ and the distortion is given by a $\mu$-distortion.

When many scatterings occur, the spectrum is driven towards kinetic equilibrium with respect to Compton scattering. Again neglecting emission and absorption processes, the kinetic equation thus becomes quasi-stationary
\beal{
\nonumber
0 \approx  \frac{\The}{x^2}\frac{\partial}{\partial x} x^4 \left[\frac{\partial}{\partial x} n +\frac{\Tz}{\Te}\,n(1+n) \right].
}
The trivial solution is $n_{\rm bb}=1/(\expf{x}-1)$ if $\Te\equiv \Tz$, since $\partial_x n_{\rm bb}=-n_{\rm bb}(1+n_{\rm bb})$. However, this is not the general solution of the problem. To find a more general solution we have to solve the equation $\partial_x n=-\frac{\Tz}{\Te}\,n(1+n)$. The factor $\Tz/\Te$ can be absorbed by redefining the frequency, $x\rightarrow \xe$, such that we now have $\partial_{\xe} n=-n(1+n)$. This can be integrated to $\ln(1+n)-\ln(n)\equiv \xe+{\rm const}$, or 
\beal{
\nonumber
n_{\rm BE}=\frac{1}{\expf{\xe+\mu_0}-1},
}
where we introduced the integration constant $\mu_0$. This is a {\it Bose-Einstein spectrum} with constant chemical potential\footnote{Notice that the sign is different from the normal convention used in thermodynamics, and the chemical potential is also defined as dimensionless parameter.} $\mu_0$. Let us pause for a moment. Photons have no rest mass, so the chemical potential should vanish, or should it not? This statement is only true if we are in full equilibrium, i.e., we have a blackbody at the temperature of the medium. More generally, for fixed photon number and energy densities the chemical potential can be non-zero. From the example we gave in Sect.~\ref{sec:thermalization is what}, we already understood that now we expect $T_\rho \neq T_N$, implying that we need at least two parameters to determine the photon distribution.

The chemical potential can in principle be both positive or negative:
\begin{itemize}
\setlength{\itemindent}{\listindentation}
\vspace{2mm}
\item $\mu_0>0$: {\it fewer} photons than in a blackbody at $\Te$ \hspace{5mm}$\rightarrow$\hspace{5mm} {\it energy release}

\item $\mu_0\equiv 0$: blackbody at temperature $\Te$ 
\qquad\;\;\,\,\quad\hspace{6.2mm} $\rightarrow$ \hspace{5mm}{\it full equilibrium}

\item $\mu_0<0$: {\it more} photons than in a blackbody at $\Te$ 
\hspace{5mm}\,$\rightarrow$ \hspace{5mm}{\it energy extraction}
\vspace{2mm}
\end{itemize}
In practice, the solution $\mu_0={\rm const} <0$ is unphysical unless $\mu_0$ is actually a function of frequency. The reason is that $\xe + \mu_0$ can vanish at $\xe=-\mu_0>0$, but this state is never reached or even passed through during the evolution, since instead excess photons would form a Bose-condensate at $x=0$ with $\mu_0=0$. In a real plasma, BR and DC emission will prevent this from happening though and in addition stimulated scattering effects become important. 
We also point out that without the term $\propto n^2$, we would have found a Wien spectrum $n_{\rm W}=\expf{\xe+\mu_0}$, which does not have these issues but also is not a blackbody spectrum. This highlights the importance of stimulated scattering processes for the formation of the CMB spectrum.

To determine the integration constant, $\mu_0$, one has to carefully compare the number and energy densities before and after the heating process. Since matter plays such a minor role in the Universe, we can then simply integrate the equilibrium equations for the photon field. Assuming that we start with a blackbody at a temperature $T_i$ and then change the energy density and number density of the photon field by some small amounts, $\Delta \rho_\gamma/\rho_\gamma$ and $\Delta N_\gamma/N_\gamma$, respectively, this consideration then leads to \citep{Sunyaev1970mu}
\beal{
\label{eq:sols_mu_DT_T_a}
\mu_0&\approx 
\frac{3}{\kappa}\left[\frac{\Delta \rho_\gamma}{\rho_\gamma}-\frac{4}{3}\frac{\Delta N_\gamma}{N_\gamma}\right]
\approx 1.401\left[\frac{\Delta \rho_\gamma}{\rho_\gamma}-\frac{4}{3}\frac{\Delta N_\gamma}{N_\gamma}\right],
&\frac{\Delta T}{T_i}
&\approx 0.6389 \,\frac{\Delta \rho_\gamma}{\rho_\gamma} - 0.5185 \,\frac{\Delta N_\gamma}{N_\gamma}
\approx 0.4561 \mu_0 + \frac{1}{3} \frac{\Delta N_\gamma}{N_\gamma}
}
with $\kappa\approx 2.1419$. The numerical coefficients are all related to exact integrals of a Planckian over frequency, but we omit the details here. From Eq.~\eqref{eq:sols_mu_DT_T_a} we see that for $\Delta \rho_\gamma/\rho_\gamma\equiv (4/3)\Delta N_\gamma/N_\gamma$ the distortion vanishes, i.e., $\mu_0=0$. This is the special case of balanced energy and photon injection, which does not occur in many distortion scenarios. For balanced injections, only the temperature of the blackbody is increased\footnote{We neglect the small heat capacity of the electrons and baryons for this statement.} after Compton scattering redistributed all photons, $\Delta T/T_i\approx \frac{1}{3} \Delta N_\gamma/N_\gamma$.

{\it But how do we define the $\mu$-distortion spectrum?} 
In the derivations from above, we used 
\beal{
\nonumber
n_{\rm BE}=\frac{1}{\expf{\xe+\mu_0}-1}\approx \frac{1}{\expf{\xe}-1}- \frac{\expf{\xe}}{(\expf{\xe}-1)^2}\mu_0+\mathcal{O}(\mu_0^2)
}
for $\mu_0\ll 1$. This suggests that $\Delta n=n_{\rm BE}(\xe)-\nbb(\xe)=-\mu_0\,\expf{\xe}/(\expf{\xe}-1)^2$ could be called the distortion with respect to the blackbody part at temperature $\Te$, and in fact this definition has been used frequently. However, since also the final electron temperature, $\Te\rightarrow T_f$, depends on $\mu_0$, this definition does not separate the distortion cleanly. Denoting the $\mu$-distortion spectrum by $M(x)$ and motivated by the fact that Compton scattering conserves photon number, one natural definition is to fix the $\mu$-distortion such that the photon number integral vanishes, i.e., $\int x^2 M(x)\id x=0$. Integrating $\Delta n$ gives $\int x^2 \Delta n \id x=-2\mu_0\int x \id x /(\expf{x}-1)=-\mu_0\,\pi^2/3 \approx -3.2899\,\mu_0$, such that 
\beal{
\label{eq:def_mu_distortion}
M(x)=\frac{x \expf{x}}{(\expf{x}-1)^2}\left[\alpha_\mu -\frac{1}{x}\right]
}
with $\alpha_\mu=\pi^2/18\zeta(3)\approx 0.4561$ fulfills the requirement $\int x^2 M(x)\id x=0$. This is the now frequently-used form for the $\mu$-distortion spectrum that we were after! And with Eq.~\eqref{eq:sols_mu_DT_T_a} we even know how to estimate the $\mu$-parameter once we know the energy and photon injection. We will return to this aspect below after one small refinement that includes the production of photons (see Sect.~\ref{sec:distortion_vis}).

\begin{figure}
\centering
\includegraphics[width=0.78\columnwidth]{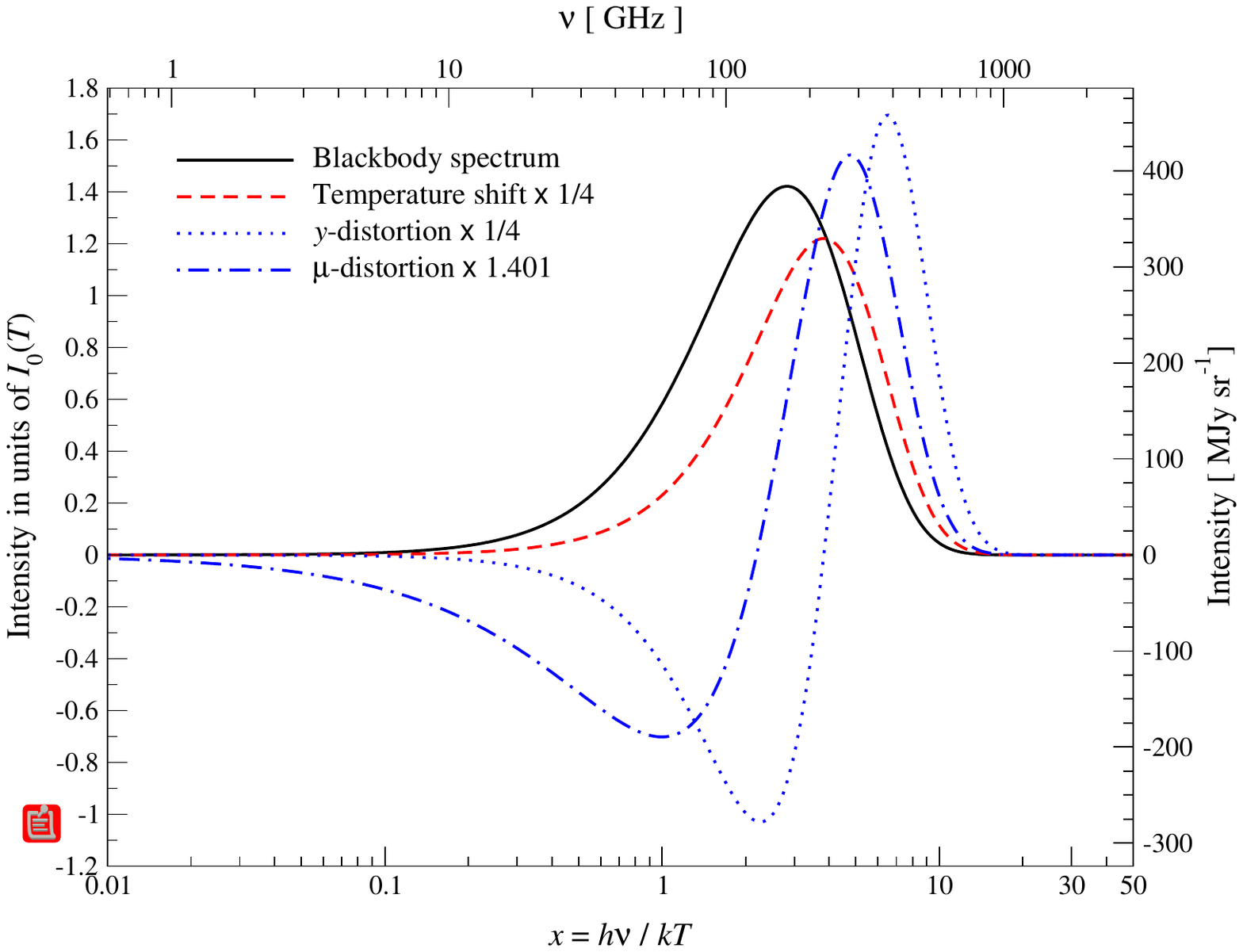}
\caption{Comparison of a Compton $y$-distortion, $Y(x)$, and $\mu$-distortion, $M(x)$, with the blackbody spectrum and temperature shift, $G(x)$. For convenience, we plot the intensity, $I_\nu=(2h\nu^3/c^2)\,n(\nu)$, as a function of $x=h\nu/kT$ and normalize the left $y$-axis by $I_0(T)=(2h/c^2)(kT/h)^3\approx 270\,\MJysr (T/2.726\Kel)^3$. The $y$-distortion has its crossover frequency around $x\simeq 3.830\,(\equiv 217\GHz)$, while the $\mu$-distortion has its zero around $x\simeq 2.192\,(\equiv 124\GHz)$. The peak of the CMB blackbody is around $x\simeq 2.821\,(\equiv 160\GHz)$, and the maximum of a temperature shift coincides with the null of the $y$-distortion. The upper $x$-axis and right $y$-axis also give the corresponding frequency and spectral intensity for $T=2.726\,\Kel$. We normalized the spectra to equal the energy density of the CMB blackbody spectrum. The distortions are therefore usually suppressed by at least four orders of magnitude. Figure is taken from \citet{Chluba2018Varenna}.}
\label{fig:Y_SZ_and_mu_distortion}
\end{figure}
\subsection{Illustration of the main distortion spectra}
\label{sec:Y_mu_dist_illustration}
We already took a big step forward in our understanding of distortion physics by learning about the two classical distortion types. In this section, we briefly illustrate these spectra. For completeness, we also introduce the spectrum of a {\it temperature shift}
\beal{
\label{eq:def_temperature_shit}
G(x)=\frac{x \expf{x}}{(\expf{x}-1)^2},
}
which is required to change a blackbody from one temperature, $T_i$, to another at $T_f=T_i(1+\Delta T/T)$ when $|\Delta T/T| \ll 1$, i.e., $\nbb(\nu, T_f)\approx \nbb(\nu, T_i)+G(h\nu/\kB T_i)\,\Delta T/T_i$. This spectrum is also relevant to the CMB temperature anisotropies and their interpretation. Indeed, the largest CMB anisotropy, which is the CMB dipole caused by our motion relative to the CMB restframe, can be readily show to have this spectrum using a Lorentz Boost of the average CMB blackbody.

In Fig.~\ref{fig:Y_SZ_and_mu_distortion}, we illustrate the frequency dependence of the temperature shift, a $y$-distortion and a $\mu$-distortion. We normalized all the spectra to equal the energy density of the CMB blackbody. The corresponding normalization factors are $N_G=1/4$, $N_Y=1/4$, and $N_M=1.401$ for each of the spectra. This makes all of the spectra comparable in amplitude, which is only for illustration. One of the crucial aspects is that the $\mu$-distortion and the $y$-distortion have differing spectra. This allows us to distinguish them with sensitive measurements. And as we have just learned, the $\mu$-distortion requires dense and sufficiently hot environments to be formed. The conditions for forming a $\mu$ distortion are unique and therefore a messenger from the early Universe. If we observe this type of distortion, we immediately know that it originated from $z\gtrsim \pot{5}{4}$, putting us into the infant Universe when it was only a few thousand years old. We will see below that there are standard sources of $\mu$-distortion in $\Lambda$CDM, making them a unique test of inflation physics and the early Universe.

\subsection{Inclusion of photon production in the $\mu$-era}
\label{sec:distortion_vis}
Although we now have a simple understanding of the main distortion types, with the late-type $y$-distortion in contrast to the early-type $\mu$-distortion, we still cannot answer when distortions become invisible. Clearly, at sufficiently early times in the $\mu$-era we expect the thermalization process to become extremely efficient and reduce the distortion amplitude to unobservable levels, i.e., $\mu\rightarrow 0$. To understand this, we also have to include the production of photons by Bremsstrahlung (BR) and double Compton (DC), which we omitted so far.

It is straightforward to approximately include the effect of BR and DC in the $\mu$-distortion era. This problem was first solved by \citet{Sunyaev1970mu}.\footnote{For a detailed discussion of the required approximations and their limitations we refer the interested reader to \citet{Chluba2014}.}
Since scattering is efficient, we can assume that the spectrum evolves along a sequence of quasi-stationary stages. However, now we also have to account for emission and absorption, such that 
\beal{
0 \approx  \frac{\The}{x^2}\frac{\partial}{\partial x} x^4 \left[\frac{\partial}{\partial x} n +\frac{\Tz}{\Te}\,n(1+n) \right] + \frac{K\,\expf{-\xe}}{x^3}\Big[1-n(\expf{\xe}-1)\Big]
}
determines the CMB spectrum. Inserting $n\approx 1/(\expf{\xe}-1)-\mu(z, \xe)\,G(\xe)/\xe$ and assuming $\xe\ll 1$ yields a simple differential equation for $\mu(z, \xe)\ll 1$, namely $0 \approx \partial_{\xe} \xe^2 \partial_{\xe} \mu - \frac{K/\Thz}{\xe^2} \mu$, which has the approximate solution 
\beal{
\label{eq:mu_x}
\mu(z, \xe)\approx \mu_0(z) \,\expf{-\xc(z)/\xe}.
}
This solution was first derived by \citet{Sunyaev1970mu}. The critical frequency, $\xc\approx \sqrt{K/\Thz}$, is determined by the competition between photon emission / absorption and Compton up-scattering of photons. Including both DC and BR, one usually has $\xc(z)\simeq 10^{-3}-10^{-2}$ during the thermalization period, which in hindsight justifies the approximations we used.

Equation~\eqref{eq:mu_x} shows that at high frequencies $\xe\gg \xc$ the chemical potential becomes constant, $\mu(z, \xe)\approx \mu_0(z)$, while at low frequencies it vanishes exponentially, returning to a blackbody at the temperature of the electrons, with a smooth transition between these extremes around $\xe\simeq \xc$. The solution has the expected limiting behavior, even if strictly speaking it is only valid at low frequencies. Most of the energy of the spectrum is at $\xe\gg \xc$, such that we already understand that $\mu_0$ can again be fixed by considering the total energetics.

With Eq.~\eqref{eq:mu_x}, one can compute the total photon production rate at any redshift. With this, one can estimate how the high-frequency photon chemical potential is affected by photon production at low frequencies. The crucial ingredient is that photons produced at low frequencies are up-scattered by Compton scattering to gradually replenish the deficit of photons, thereby diminishing the amplitude of the distortion. 
The analytic treatment essentially boils down to a differential equation for $\mu_0(z)$, which for single energy release, $\Delta \rho_\gamma/\rho_\gamma|_i$, at redshift $z_i$ has the solution
\vspace{-3mm}
\beal{
\label{eq:mu_sol_sing}
\mu_0(z)
&\approx 
1.401\left.\frac{\Delta\rho_\gamma}{\rho_\gamma} \right|_i
\,\Jbb(z_i, z),
}
where $\Jbb(z_i, z)$ define the {\it spectral distortion visibility function} between the injection redshift $z_i$ and $z$ (with $z<z_i$). It determines the fraction of energy injected at $z_i$ that is still visible as a distortion at redshift $z$. Including only DC emission, one has 
\beal{
\Jbb(z_i, z)\approx \JDC(z_i, z) = \expf{-(z_i/z_\mu)^{5/2}+(z/z_\mu)^{5/2}}
}
with thermalization redshift \citep{Sunyaev1970mu, Danese1982, Burigana1991, Hu1993}
\beal{
\zmu\approx\zmudc\approx \pot{1.98}{6} \left[\frac{\Omega_{\rm b} h^2}{0.022}\right]^{-2/5} \left[\frac{T_0}{2.725 \Kel}\right]^{1/5}
\left[\frac{(1-\Yp/2)}{0.88}\right]^{-2/5},
}
assuming $N_{\rm eff}=3.046$. For $\Jbb(z_i, z)\simeq 1$, most of the injected energy is still stored in the distortion, while for $\Jbb(z_i, z)\ll 1$, most of the energy was thermalized and converted into a temperature shift, with a spectrum related to $G(x)$. 
This implies that after a single energy release event, today's remaining chemical potential is heavily suppressed if the energy injection happens at redshifts $z\gtrsim z_{\rm dc}\approx \pot{1.98}{6}$ or before about $\simeq 3$~months after the Big Bang. 
This is a remarkable result: it essentially takes the whole age of the Universe to form the CMB blackbody spectrum we see today! If we inject some energy at 13.7 Billion years minus 3 months, we will see a distortion. The main catalyst for photon production and up-scattering (the baryons) is simply too rare.

Moving from single energy release to continuous energy release, we can then estimate the final $\mu$-distortion measured today ($z=0$) by summing the contributions from individual injections weighted by the distortion visibility function. This gives the simple approximation  
\beal{
\label{eq:mu_improved}
\mu_0 &\approx 1.401 \left.\frac{\Delta \rho_\gamma}{\rho_\gamma}\right|_\mu 
\approx 1.401 \int_{z_{\mu y}}^\infty \frac{\id ({Q}/\rho_\gamma)}{\id z'}\,\Jbb(z', 0) \id z',
}
where $\id ({Q}/\rho_\gamma)/\id z$ describes the energy release relative to the CMB blackbody and depends on the specific energy release mechanism (see Sect.~\ref{sec:mechanisms}). We neglected any external photon production, which can also be analytically treated \citep{Chluba2015GreensII}.

{\it Importance of DC over BR} -- In the treatment summarized above, the DC process dominated in the definition of the thermalization redshift. At $z\gg\zmudc $, thermalization is very efficient and the distortion visibility drops exponentially. If alternatively we only include BR emission, we find \citep{Sunyaev1970mu, Danese1982, Hu1993, Chluba2014}
\beal{
\Jbb(\zh, 0)\approx \mathcal{J}_{\rm BR}(\zh)=\exp\left(-[\zh/z_{\rm br}]^{1.328}\right)
}
with $z_{\rm br}\approx \pot{5.27}{6}$. This shows that the thermalization redshift is significantly higher when only BR is included. In addition, the distortion visibility function drops less rapidly, making the CMB a lot more vulnerable to distortion, and highlighting how crucial the DC process is.

\subsection{Simplest description of primordial distortions}
\label{sec:zeroth_order_picture}
We now have all the pieces for a simplest, zeroth order description of primordial distortion signals created by energy injection. No matter when in the history of the Universe find ourselves, energy release momentarily always creates $y$-type distortions as long as the scattering $y$-parameter is small. Compton scattering then gradually converts the $y$-distortion spectrum into a $\mu$ distortion. If the injection happened at $z\lesssim z_{\mu y}=\pot{5}{4}$, this conversion remains incomplete and we can simply label the distortion as a $y$-distortion. Since in this era, the up-scattering of soft photons produced by BR is inefficient, almost all the injected energy is found in the distortion, i.e., the distortion visibility is $\Jbb(z, 0)\approx 1$. If the injection happens at $z\gtrsim z_{\mu y}$, we can approximate the distortion as a $\mu$-distortion. However, the distortion amplitude can be diminished by the up-scattering of low-frequency photons, which thermalize the high-frequency spectrum according to the distortion visibility function, $\mathcal{J}_\mu(z)\approx \JDC(z, 0)=\expf{-(z/\zmu)^{2.5}}\leq 1$. For every injection, the fraction of energy that is visible as a $\mu$-distortion is given by $\mathcal{J}_\mu(z_{\rm i})$ and the corresponding fraction of energy that was thermalized then is $\mathcal{J}_T(z_{\rm i})=1-\mathcal{J}_\mu(z_{\rm i})$. 

With this simple picture in mind, the corresponding $y$, $\mu$ and temperature parameters can be estimated as
\beal{
\label{eq:ymuT_improved}
y &\approx \frac{1}{4} \int^{z_{\mu y}}_0 \frac{\id ({Q}/\rho_\gamma)}{\id z'} \id z', 
&\mu &\approx 1.401 \int_{z_{\mu y}}^\infty \frac{\id ({Q}/\rho_\gamma)}{\id z'}\,\mathcal{J}_\mu(z') \id z',
&\frac{\Delta T}{T} &\approx \frac{1}{4}  \int_{z_{\mu y}}^\infty \frac{\id ({Q}/\rho_\gamma)}{\id z'}[1-\mathcal{J}_\mu(z')] \id z',
}
by just considering the energetics of the problem. This leading order description is extremely elegant and simple, and captures the main features of the cosmological thermalization problem well. From \COBEF, we have the distortion limits $|y|\lesssim \pot{1.5}{-5}$ (95\% c.l.) and $|\mu|\lesssim \pot{9.0}{-5}$ (95\% c.l.) \citep{Fixsen1996}, obtained some 30 years ago. These imply that energy release at $z\lesssim \pot{2}{6}$ is restricted to\footnote{This statement follows by converting $\Delta \rho_\gamma/\rho_\gamma=4 y$ and $\Delta \rho_\gamma/\rho_\gamma=\mu/1.401$.} $|\Delta \rho_\gamma/\rho_\gamma| \lesssim \pot{6}{-5}$ (95\% c.l.), ruling out cosmologies with extended energy injection. However, as we shall see below, even in $\Lambda$CDM we expect energy release below this long-standing limit from \COBEF, promising a detection of distortions in the near future.

\subsection{New epoch-dependent information beyond the leading order picture}
\label{sec:Transition}
Even if the simple picture presented in the previous sections is extremely powerful, it misses some important aspects that make SD science a lot more exciting. Importantly, the transition between $\mu$ and $y$ is not abrupt at $z\simeq \pot{5}{4}$ but happens gradually over a range of redshifts, where in the transition regime the distortion is not only given by the superposition of $\mu$ and $y$-distortion. This makes the distortion signal much richer, with new epoch-dependent information that provides extra leverage on distortion scenarios \citep[e.g.,][]{Hu1995PhD, Chluba2011therm, Khatri2012mix, Chluba2013Green}. Because the distortions are assumed to be small, the full physics can be captured by computing the Green's function of the problem, $G_{\rm th}(\nu, z)$. The distortion from energy injection is then given by
\beal{
\label{eq:Greens}
\Delta I_\nu\approx 
\int_0^\infty G_{\rm th}(\nu, z')\frac{\id ({Q}/\rho_\gamma)}{\id z'} \id z'.
}
Cast into the Green's function language, the description presented in Sect.~\ref{sec:zeroth_order_picture} can be summarized as\footnote{Here we use $x=h\nu/\kB T_0=1.76\,[\nu/100\GHz]$.}
\citep{Chluba2013Green}
\beal{
\label{eq:Greens_simplest}
G_{\rm th}(\nu, z)\approx 
\frac{2h \nu^3}{c^2}\times 
\begin{cases}
\frac{Y(x)}{4} &\qquad \text{for $z\leq z_{\mu y}$}
\\[2mm]
1.401 M(x) \, \mathcal{J}(z) + \frac{G(x)}{4} [1-\mathcal{J}(z)] &\qquad \text{for $z> z_{\mu y}$},
\end{cases}
}
although this does not capture any of the new aspects yet. Analytically it is indeed difficult to go beyond; however, we can easily compute the thermalization Green's function numerically using {\tt CosmoTherm}, yielding a quasi-exact representation of the problem. The procedure is very simple: we compute the distortion response for single energy injections at varying injection redshifts, evolving the CMB spectrum until $z=0$. The resultant signal at $z=0$ is the desired Green's function, $G_{\rm th}(\nu, z)$, which can be tabulated and then interpolated.

\begin{figure} 
   \centering
   \includegraphics[width=0.8\columnwidth]{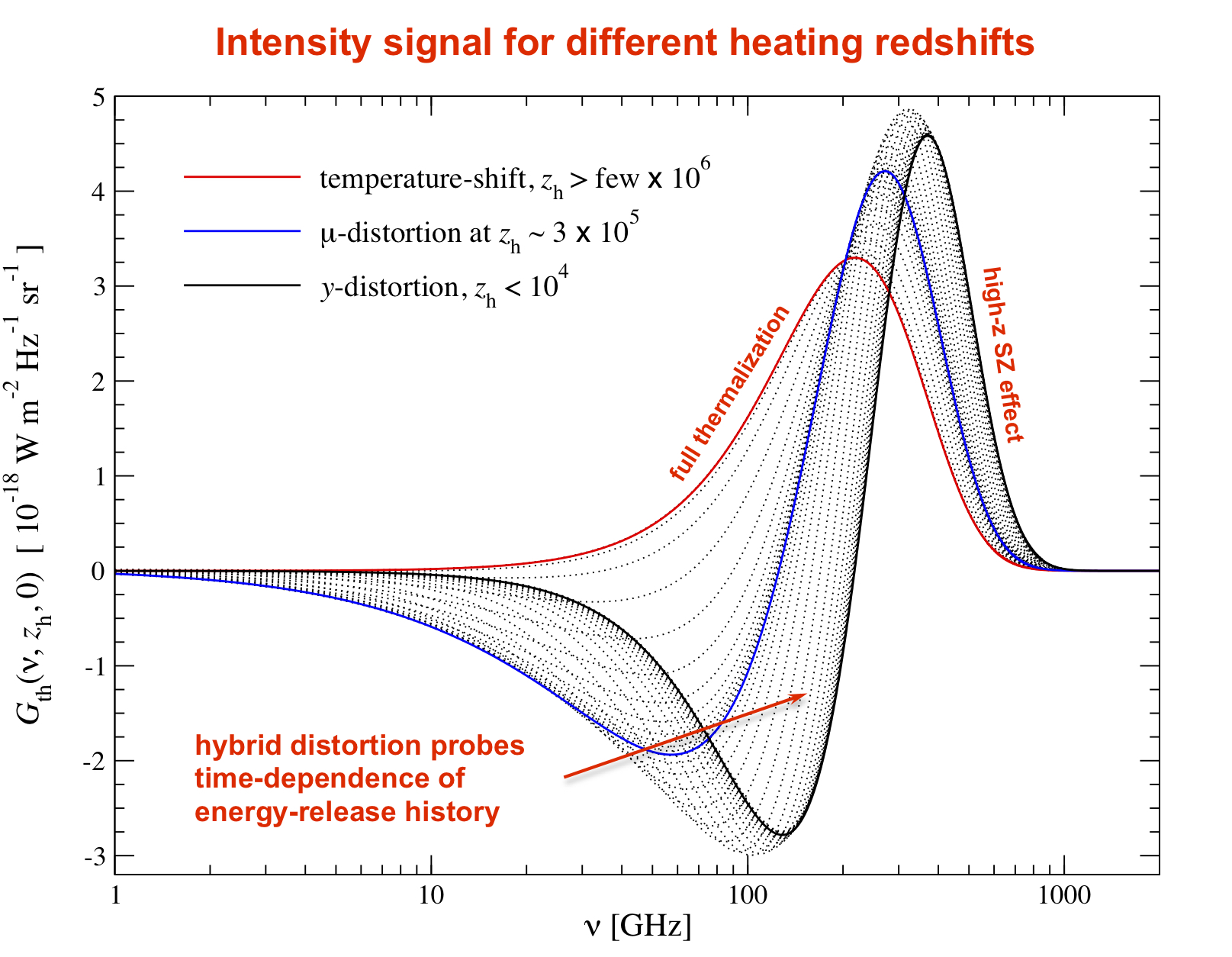}
   \\
   \includegraphics[width=0.8\columnwidth]{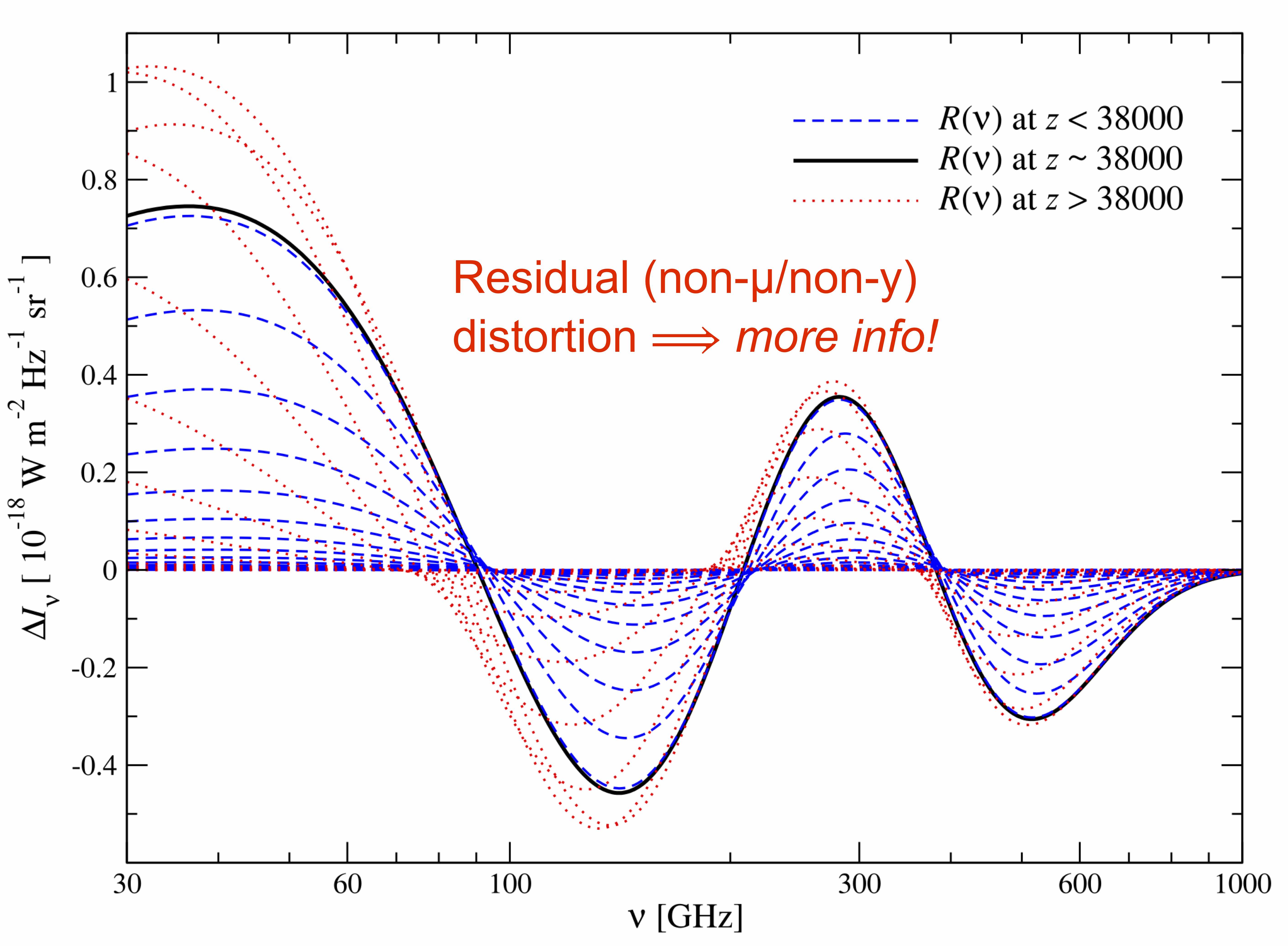}
   \caption{Change in the CMB spectrum after a single energy release at different heating redshifts, $z_{\rm h}$. Top panel: thermalization Green's function, $G_{\rm th}(\nu, z_{\rm h})$ -- lower panel: residual distortion. At $z\gtrsim \pot{\rm few}{6}$, a temperature shift is created. Around $z \simeq \pot{3}{5}$ a pure $\mu$-distortion appears, while at $z\lesssim 10^4$ a pure $y$-distortion is formed. At all intermediate stages, the signal is given by a superposition of these extreme cases with a small residual (non-$\mu$/non-$y$) distortion that contains information about the time-dependence of the energy-release process. Figures is adapted from \cite{Chluba2013Green} and \cite{Chluba2013PCA}.}
   \label{fig:Greens}
\end{figure}

The numerical results for $G_{\rm th}(\nu, z)$ are illustrated in Fig.~\ref{fig:Greens}. We can see that for heating redshifts $\zh\lesssim 10^4$, the Green's function is essentially that of a $y$-distortion, $\propto Y(x)$. For $\zh\simeq \pot{3}{5}$, we recover the $\mu$-distortion spectrum, $\propto M(x)$, while for $\zh\gtrsim \pot{\rm few}{6}$ we obtain a temperature shift, $\propto G(x)$, with subdominant (yet always present) $\mu$-distortion. All other injection cases smoothly transition between these extremes, giving a complete picture for the distortion signal when used in Eq.~\eqref{eq:Greens}. Since the Green's function can be computed once, this allows us to study various distortion scenarios very efficiently.

However, the {\it crucial} aspect now is that the intermediate cases are {\it not} just given by the sum of $G(x), Y(x)$ and $M(x)$. 
By eliminating the leading order $\Delta T/T$, $\mu$ and $y$-distortion contributions\footnote{To remove $\Delta T/T$, a photon number integral constraint is imposed to cleanly isolate the distortion.}, one is left with a smaller signal, the so called {\it residual distortion} or $r$-type distortion (see bottom panel of Fig.~\ref{fig:Greens}), which can be conveniently parametrized using distortion eigenmodes \citep{Chluba2013PCA}. The $r$-type distortion is what contains the extra epoch-dependent information and detection limits for various energy release scenarios were, for instance, presented in \citet{Chluba2013PCA}. For sufficiently large distortions ($\leftrightarrow$ precise distortion measurements), we may be able to access this valuable information to help distinguish different scenarios with future CMB spectrometers.

We comment that slightly improved versions of the analytic Green's function in terms of $\Delta T/T$, $\mu$ and $y$ can be obtained by studying the full numerical results. Some frequently-used approaches are compared in \citet{Chluba2016}, highlighting the precision of each scheme. In addition, a new frequency discretization [with a basis defined by applying the boost operator $\mathcal{\hat{O}}_z=-x\partial_x$ multiple times to $Y(x)$] was recently developed and shown to almost exactly represent the full thermalization Green's function \citep{Chluba2023FHI}. In this framework, only $17$ distortion parameters have to be evolved across redshift, eliminating the necessity for using pre-computed results in distortion analyses. However, for more accurate computations, the pre-tabulated Green's function results can always be applied reliably. 

\begin{figure}
\centering
\includegraphics[width=0.9\columnwidth]{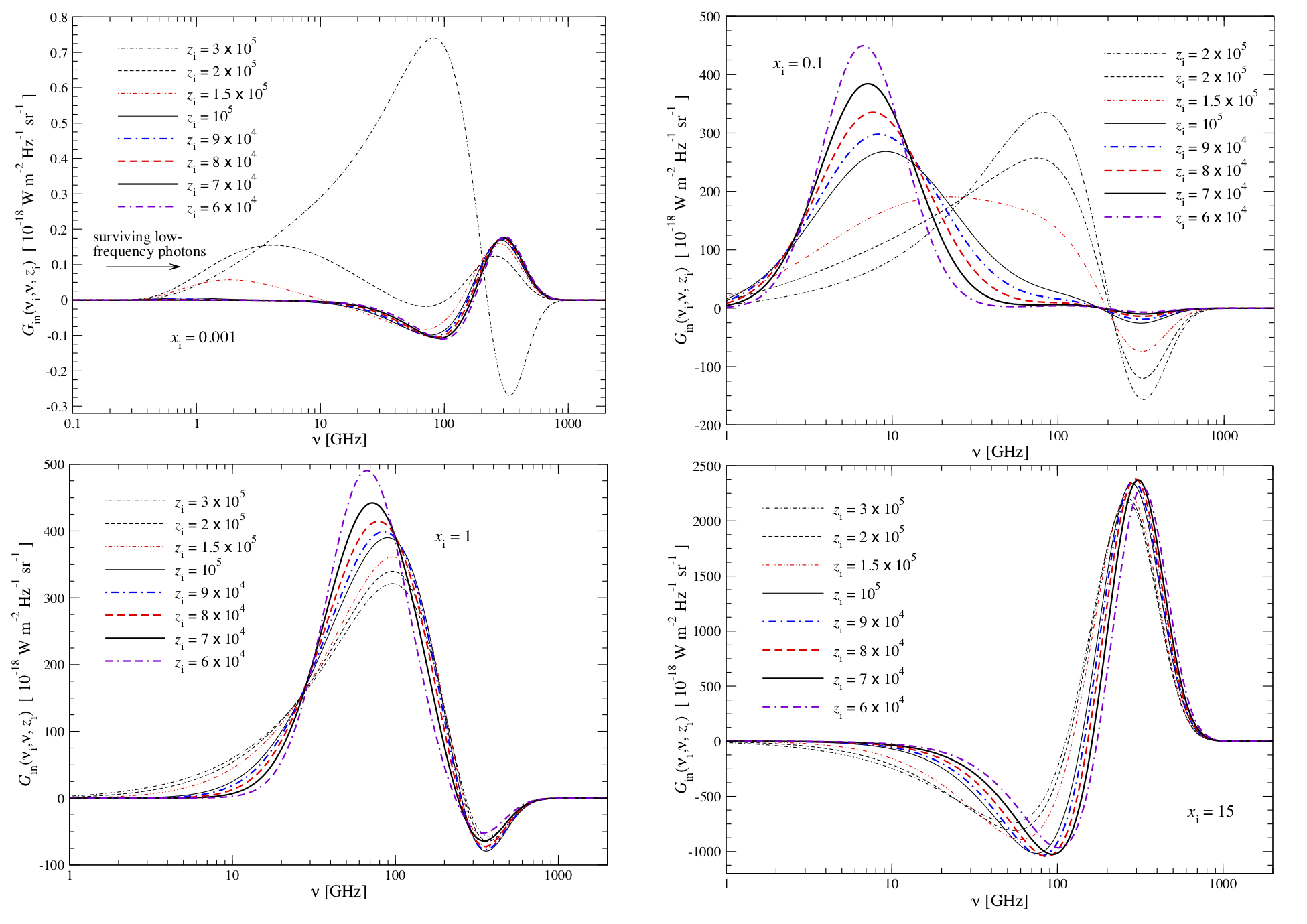}
\caption{Spectral distortions created by photon injection at different frequencies and initial redshifts. A rich phenomenology of distortion shapes can be created from processes with photon injection, in principle helping us to further distinguish different scenarios using CMB spectroscopy. The Figure is taken from \cite{Chluba2015GreensII}.}
\label{fig:photon_inj_dist}
\end{figure}

\subsection{Additional aspects of distortion physics}
\label{sec:additional_SD_physics}
The distortion physics discussed in the previous sections assumed that the signals are created by energy injection processes that lead to {\it heating} of the electrons/baryons. Here, we now briefly mention distortions created by photon injection, non-thermal electron populations and when the distortions are large, at least for part of their evolution. Together these generalizations create additional ways to learn about the various distortion scenarios, going beyond the simple thermalization Green's function of the previous section.

\subsubsection{Distortions from photon injection}
\label{sec:photon_injection}
Distortions created by photon injection processes can exhibit a much richer phenomenology than the broad $\mu$-/$y$-distortion spectra created by energy release. This is illustrated in Fig.~\ref{fig:photon_inj_dist} for several cases, showing that the final distortion depends on both the injection time and frequency.
In terms of physics, distortions created by photon injection do not directly heat the electrons or baryons but add energy and entropy to the photon field. Only once Comptonization becomes relevant (i.e., the scattering $y$-parameter $y_{\rm sc}=\int \Thz \id \tau$ becomes sizeable) do the electrons start to heat or cool. The net effect depends on the injection frequency of the photons. For injections at $x_i =h\nu_i/\kB \Tz \gtrsim 3.6-3.8$, photons on average loose energy and heat the baryons. In return, this causes a broad $\mu$- and $y$-type contribution to the total distortion signal, which for high frequency injection, $x_i \gtrsim 10$, can dominate. At lower frequencies, the electrons cool since photons up-scatter on average. This can create negative $\mu$ and $y$-type contributions, as the required energy is taken from the CMB.

Photon injection distortions are by no means exotic. For example, the cosmological recombination radiation \citep{Sunyaev2009} -- one of the standard $\Lambda$CDM distortions -- is created primarily by photon injection. Injection of photons can also occur in decaying or annihilating particle scenarios or evaporation of primordial black holes. In light of the measurements of EDGES \citep{Edges2018} and the ARCADE low-frequency radio excess \citep{Fixsen2011, DT2018, Singal2022}, photon injection distortions of the CMB could become extremely interesting \citep{Feng2018}. This is because these observations potentially point towards a connection with photon injection (or absorption) from decaying or annihilating particles and their low energy by-products in the form of non-thermal Bremsstrahlung or synchrotron emission, although for a detailed treatment the soft photon heating on the global 21cm signal has to be carefully considered \citep{Acharya2023SPH, Cyr2024SPH}. This highlights a powerful link between CMB SD science and 21cm cosmology, which we have only started to explore.

In addition, the thermalization problem is even richer when including the effect of pre-recombination ($z\gtrsim 10^3$) atomic transitions \citep{Liubarskii83, Chluba2008c}, which ultimately are also linked to photon injection distortions. As illustrated in \citet{Chluba2008c} and also \citet{Chluba2010}, uncompensated cycles of atomic transitions can lead to extra CMB SD features that might allow us to gain deeper insight into the $\mu$- and $y$-eras. Since the strength and shape of these features depend on when the distortion is created, this adds additional epoch-dependent information that may even allow separating pre- from post-recombination energy release. Some recent consideration in the context of high-energy particle cascades can be found in \citet{Liu2023}, but a lot more work is needed.

\subsubsection{Non-thermal electrons and high energy photons}
\label{sec:non-thermal}
One of the key assumptions for the cosmological thermalization problem is that the electrons remain thermal and in equilibrium with the baryons at a common temperature (see Sect.~\ref{sec:conditions}). The relaxation of electrons and baryons towards equilibrium is determined by Coulomb interactions, which can become quite slow for high energy particles above a few keV \citep[e.g.,][]{Shull1985, Slatyer2015}. Similarly, for high energy photons, Compton scattering is no longer well-described by the Kompaneets equation, since Klein-Nishina corrections and strong recoil effects become important, necessitating a Compton scattering kernel approach \citep{Sarkar2019}. In particular for distortion scenarios from decaying or annihilating high mass DM particles \citep[e.g.,][]{Chen2004, Slatyer2009, Huetsi2009}, this can lead to non-thermal populations of electrons and high energy photons that cause non-thermal CMB distortion signals \citep[e.g.,][]{Ensslin2000, Slatyer2015, Acharya2018}. 

One of the most important aspect is that high-energy non-thermal electrons can up-scatter CMB photons far beyond the CMB regime, essentially removing them from the CMB bands. The high-energy photons then create secondary non-thermal electrons which further result in more high energy photons, sourcing an electromagnetic cascade. The related CMB distortion signal can thus be modified and contains additional information about the underlying particle physics scenario. This opens the exciting opportunity for more accurately constraining DM physics, should decays or annihilations that induce high energy particle cascades occur.

\subsubsection{The large distortion regime}
\label{sec:large_distortion}
Although we know from \COBEF that today the distortions are small this statement {\it does not} have to hold for all stages in evolution of the distortion. Firstly, deep into the temperature era, thermalization is extremely efficient, and the initial distortion can generally be large without violating today's constraints. Secondly, for scenarios with localized energy injection, one can also encounter large distortions, even if averaged over the full sky the distortion is indeed small. 

Secondly, it has been shown that for energy injection scenarios the thermalization process takes longer when large distortions are involved  \citep{Chluba2020large}. This is because the distortion shape in the $\mu$-era significantly departs from the simple approximation obtained in the small distortion limit. This modifies the photon production efficiency and also how quickly photons up-scatter to the high frequency domain. Indeed, these effects alter the distortion constraints for a range of scenarios in particular with energy release at $z\gtrsim \pot{2}{6}$, aspects that we are now able to model using {\tt CosmoTherm}.

\section{Observational outlook and perspectives}
\label{sec:observations}
The seminal measurements of the CMB blackbody spectrum by \COBEF in the early '90s cemented the Hot Big Bang model by ruling out any energy release greater than $\Delta \rho_\gamma / \rho_\gamma~\simeq~6~\times~10^{-5}$ (95\% c.l.) of the energy in CMB photons \citep{Mather1994, Fixsen1996}. Attempts to follow-up on \COBEF from the ground with TRIS \citep{tris1, tris2} and with the balloon-bourne ARCADE experiement \citep[e.g.,][]{arcade2}, led to no major advances, although a recent reanalysis of \COBEF data with modern methods has resulted in some improvements and additional robustness tests \citep{Bianchini2022}.

\begin{figure} 
\centering
\includegraphics[width=0.82\columnwidth]{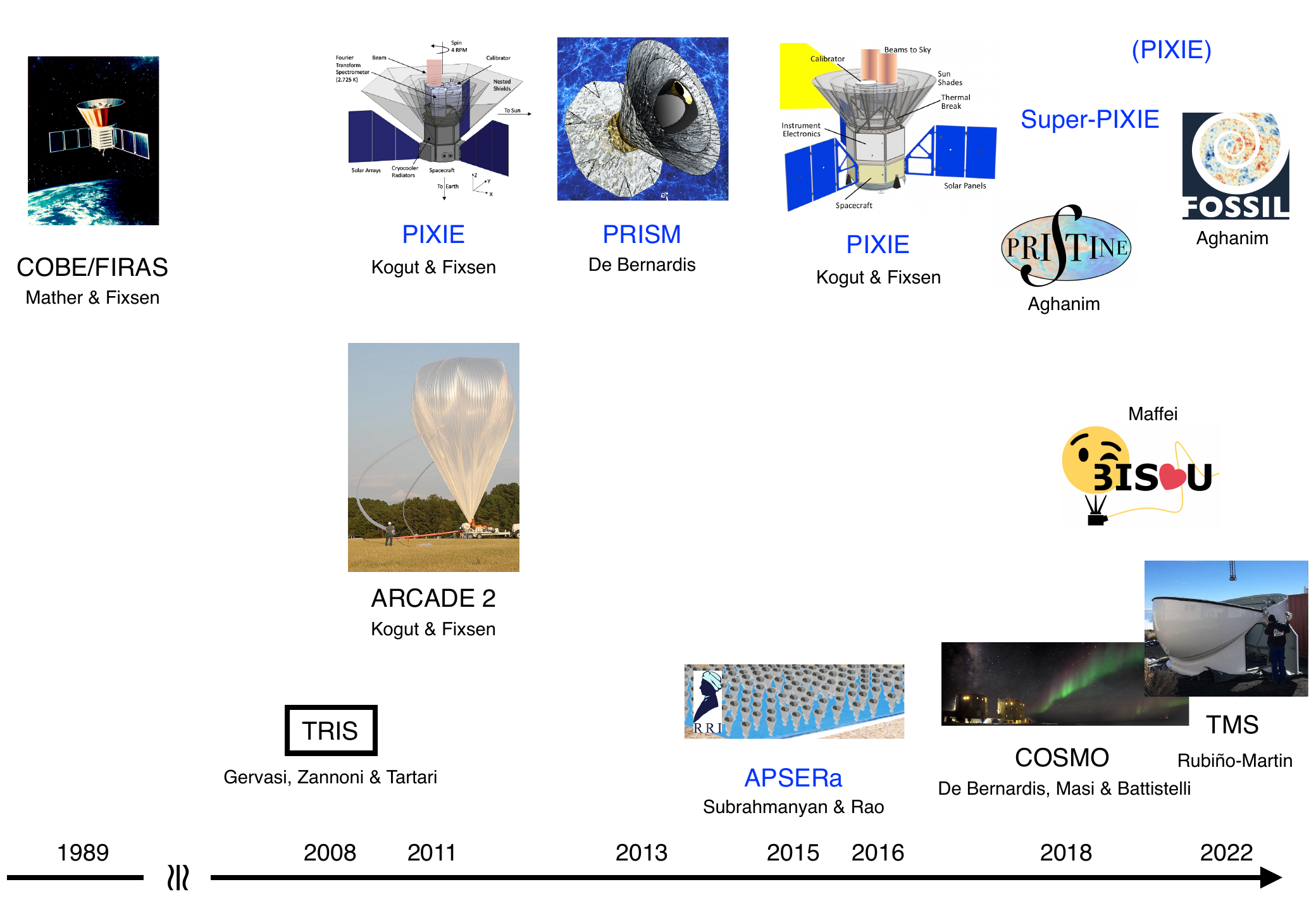}
   \caption{Some of the past CMB spectrometer experiments and proposals, in addition to ongoing efforts. The long-standing distortion limits from \COBEF  still define the state-of-the-art even if an improvement by at least three orders of magnitudes in sensitivity over \COBEF should be possible with current day-technology. APSERa, COSMO, TMS and BISOU are currently in preparation as part of the sub-orbital pathfinding activities, with the most recent news that BISOU has now entered Phases A of its development.}
   \label{fig:Missions-past}
\end{figure}

However, advances since \COBEF, in both detector technology and cryogenics, promise to improve existing constraints by three to four orders of magnitude or more [e.g., with CMB spectrometer concepts like \PIXIE \citep{Kogut2011PIXIE, Kogut2016SPIE}, \PRISM \citep{PRISM2013WPII}, \PRISTINE and {\it FOSSIL}], thereby opening an enormous discovery space for both guaranteed $\Lambda$CDM distortion signals and those caused by new physics (see Sect.~\ref{sec:mechanisms}). In the meantime, significant progress can be made from the ground with CMB spectrometers such as APSERa \citep{Mayuri2015}, COSMO \citep{Masi2021} and TMS \citep{Jose2020TMS}, the latter expecting first data on a years timescale. Additional important opportunities open up with next generation balloon-bourne experiments like BISOU \citep{BISOU2021}, which was recently confirmed to enter Phase A of its depelopment by the french space agency (CNES). Taken together, all these activities give an optimistic outlook on future CMB spectrometer experiments (see Fig.~\ref{fig:Missions-past}), with the ultimate goal to prepare for a future CMB spectrometer mission possibly within the Voyage 2050 program of ESA \citep{Chluba2021Voyage}.

Let us take a more careful look at some of the technological aspects. \COBEF was based on a so-called Fourier Transform Spectrometer (FTS) with an external calibrator source to perform absolute measurements of the sky flux. 
The sensitivity of \COBEF was not background limited; its sensitivity was instead set by {\it phonon noise} from the 1.4 K detector. Modern detectors, operating at $\simeq$~0.1~K, would have detector (dark) noise well below the intrinsic limit set by photon arrival statistics. The sensitivity of a background-limited instrument could be further improved by increasing its throughput or the total integration time and, in a less trivial way, by modifying the mirror stroke (i.e., frequency-sampling) of the FTS and reducing the optical load at high frequencies. By combining replicas of the same telescope design one can additionally enhance the sensitivity, although this can quickly exhaust budgetary constraints. Modern blackbody calibrators now also reach sufficient thermal and spectral stability for the task \citep[see][for discussion of systematic error mitigation]{Kogut2023}. 
We can therefore expect that with future FTS approaches that capitalize on the legacy of \COBEF we can in principle reach unprecedented sensitivities and spectral coverage to search for small distortion signals well below the levels of \COBEF. Significant progress can be made from the ground and balloon, possibly yielding $\simeq 20-30$ times higher sensitivity than \COBEF, being limited primarily by atmospheric residuals \citep{BISOU2021}. To reach beyond this, a dedicated CMB spectrometer mission will be needed, allowing us to dream of directly detecting some of the smallest signals from the primordial plasma. 

\begin{figure}
\centering
\includegraphics[width=0.81\columnwidth]{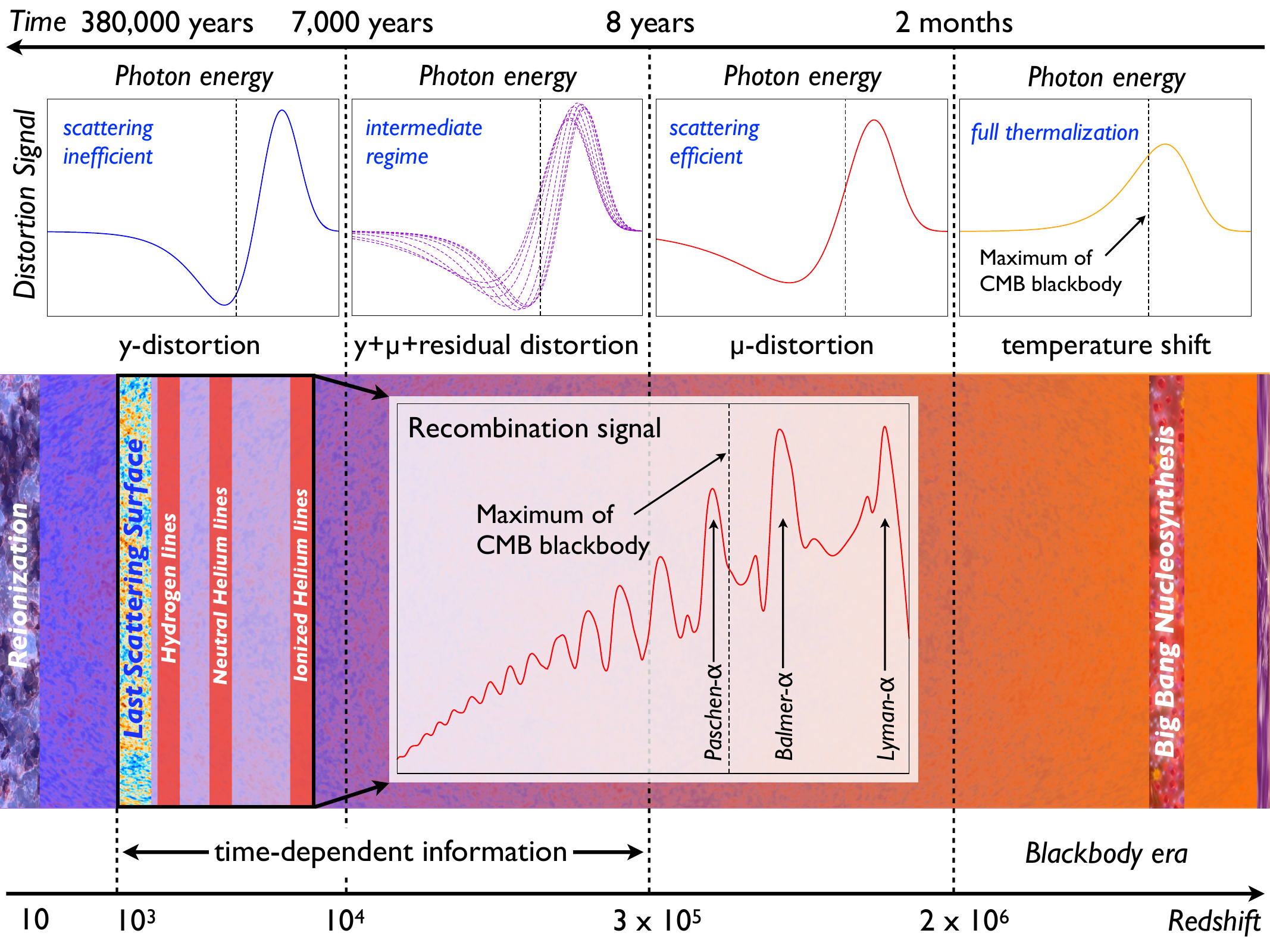}
   \caption{Evolution of spectral distortions across time. Distortions probe the thermal history over long periods deep into the primordial Universe that are inaccessible by other means. The distortion shape contains valuable epoch-dependent information that allows distinguishing different sources of distortions. Line-emission created during the cosmological recombination eras leave a detailed 'fingerprint' of the recombination process and further epoch-dependent information. The figure is adapted from \citet{Silk2014Sci}.}
   \label{fig:stages}
\end{figure}

In addition to the technological challenges, mitigation of galactic and extra-galactic foregrounds are a major concern.
A robust detection of spectral distortion signals in the presence of bright astrophysical foregrounds requires observations over multiple decades in frequency, between $\simeq$~10~GHz and a few$\times$THz. Our current understanding of the intensity foregrounds comes primarily from \Planck, \WMAP and assorted ground-based experiments. At the sensitivities of these observations, the intensity foregrounds could be modeled with sufficient accuracy using a limited set of parameters \citep[e.g.,][]{Vince2015, abitbol_pixie}. 
At high frequencies, the foregrounds are dominated by dust emission from the Milky Way and the cosmic infrared background, while at low frequencies Galactic synchrotron and free-free emission dominate. FTS concepts are well suited for observations at $\nu \gtrsim 100\,\GHz$; however, the multi-moded nature of the concept as well as simple restrictions in terms of the possible detector sizes make observations at $\nu \lesssim 30\,\GHz$ quite challenging. It was demonstrated that this is precisely where a lot of leverage for the detection of $\mu$-distortions comes from \citep{abitbol_pixie, Chluba2021Voyage}, demanding a discussion of new approaches. One possibility is to consider traditional CMB imaging technology to increase the number of detectors per frequency and optimize the distribution of channels \citep{SPECTER2024}; however, control of systematics and calibration may pose a significant challenge. Additional gains may stem from combining multiple experiments to perform a multi-tracer and multi-messenger subtraction of the low-redshift Universe in a highly synergistic future analysis framework. Indeed the CMB community is already moving towards these kind of opportunities with initiatives such as {\tt CosmoGlobe}\footnote{\url{https://www.cosmoglobe.uio.no/}}, but more work is required to explore the potential of these activities with regards to CMB SD science.

\begin{figure}
\centering 
\includegraphics[width=0.8\columnwidth]{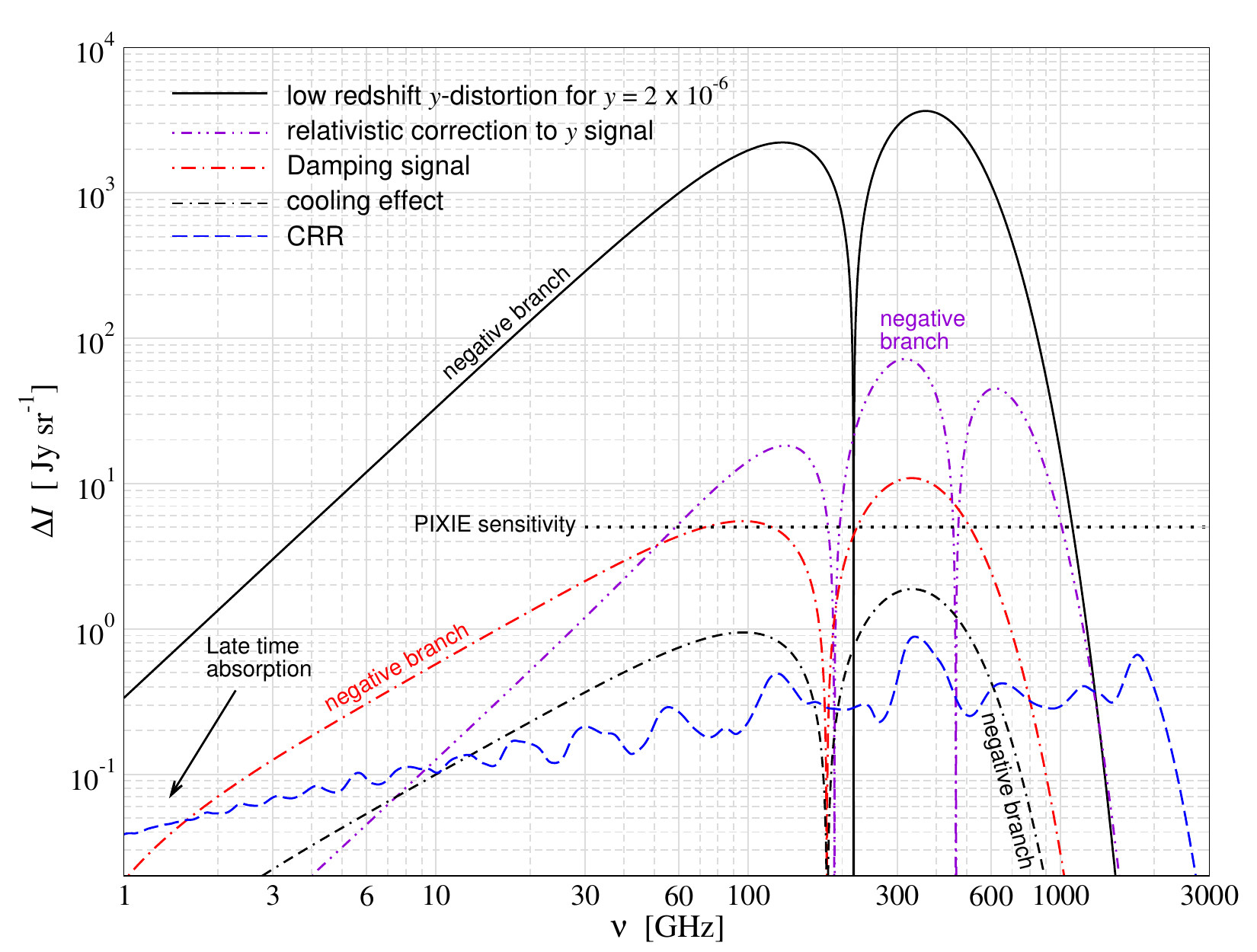}
\caption{Comparison of several CMB monopole distortion signals produced in the standard $\Lambda$CDM cosmology. 
The low-redshift distortion created by reionization and structure formation is close to a pure Compton-$y$ distortion with $y\simeq \pot{2}{-6}$. Contributions from the hot gas in low mass haloes give rise to a noticeable relativistic temperature correction, which is taken from \citet{Hill2015}. The damping and adiabatic cooling signals were explicitly computed using {\tt CosmoTherm}, providing one of the key targets to test inflation physics. The cosmological recombination radiation (CRR) was obtained with {\tt CosmoSpec} for the standard recombination history. The estimated sensitivity ($\Delta I_\nu \approx 5\, {\rm Jy/sr}$) of a \PIXIE-type experiment is shown for comparison (dotted line). The figure was taken from \cite{Chluba2016}.}
\label{fig:signals}
\end{figure}

\section{CMB spectral distortion signals from various scenarios}
\label{sec:mechanisms}
Now that we have a detailed understanding of how distortions are created and what the future observational prospects are, we shall present a discussion of various signals one could expect and what they could teach us about the Universe. 
Several exhaustive reviews exist \citep{Chluba2011therm, Sunyaev2013, Tashiro2014, deZotti2015, Lucca2020}, covering both standard and non-standard examples and some of the distortion physics. Here we highlight some of the main distortion signals expected within $\Lambda$CDM and only briefly mention more exotic sources of distortions. The different distortion stages as explained in the previous sections, are summarized again in Fig.~\ref{fig:stages} for reference. A summary of the relevant $\Lambda$CDM distortions is shown in Fig.~\ref{fig:signals}, as taken from \citet{Chluba2016}. The distortion templates are available at \url{www.chluba.de/CosmoTherm}.

\subsection{Reionization and structure formation}
\label{sect:reion}
The first sources of radiation during reionization, supernova feedback and structure formation shocks heat the intergalactic medium at low redshifts ($z\lesssim 10$), producing hot electrons (in a wide range of temperatures $\Te\simeq 10^4\,{\rm K}-10^7\,{\rm K}$) that partially up-scatter CMB photons, causing a Compton $y$-distortion. 
Although this gives rise to the {\it largest} expected average distortion of the CMB caused within $\Lambda$CDM, its amplitude is quite uncertain and depends on the detailed structure and temperature of the medium, as well as scaling relations (e.g., between halo mass and temperature).
Several estimates for this contribution were obtained, yielding values for the total $y$-parameter at the level $y\simeq \pot{\rm few}{-6}$ \citep[e.g.,][]{Refregier2000, Zhang2004, Hill2015}. 

Following  \citet{Hill2015}, we use a fiducial value of $y=\pot{2}{-6}$ (see Fig.~\ref{fig:signals}). Most of this value comes from electrons in low-mass haloes ($M\simeq 10^{13}\,M_\odot$) and the signal should be detectable with a \PIXIE-type experiment at more than $10^2\,\sigma$, and BISOU should already be capable of providing a first $5\sigma$ detection. Future CMB imagers furthermore have the potential to separate the spatially-varying signature caused by the warm hot intergalactic medium and proto-clusters, if challenges of accurate channel inter-calibration can be overcome. Cross-correlation techniques may further allow extracting tomographic information about this signal.

Because the average $y$ signal is so easily detectable with a \PIXIE-type mission, small corrections due to the high gas temperature ($k\Te \simeq 1\,\keV$) become noticeable \citep{Hill2015}. The relativistic correction can be computed using {\tt SZpack} and differs from the distortions produced in the early Universe. This correction should be detectable with \PIXIE at $\simeq 10-20\,\sigma$ and could teach us about the average temperature of the intergalactic medium and feedback processes \citep{Thiele2022}. Both distortion signals are illustrated in Fig.~\ref{fig:signals}.

\begin{figure} 
   \centering
   \includegraphics[width=0.92\columnwidth]{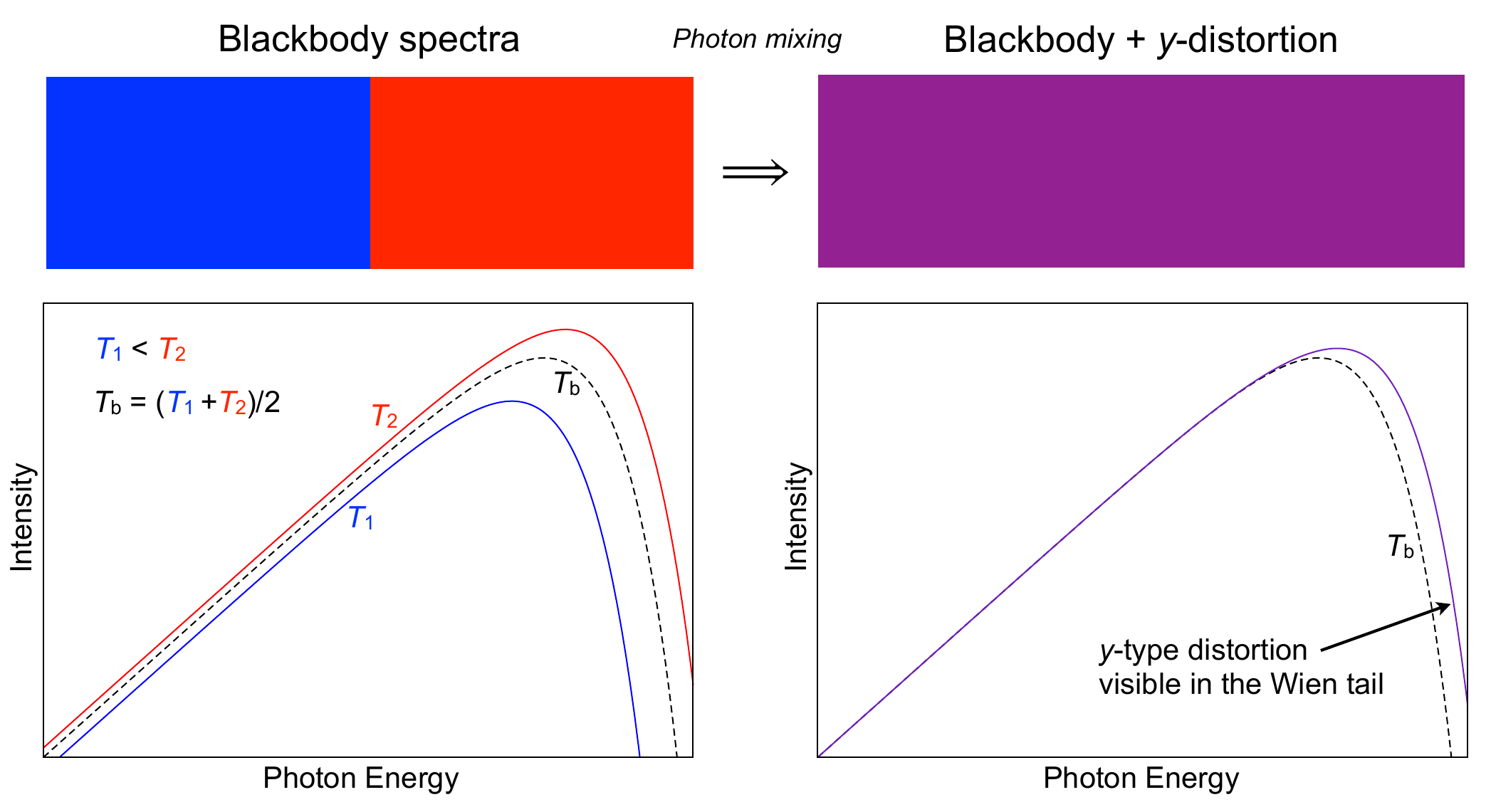}
   \caption{Illustration for the distortion created by the superposition of blackbodies. We envision blackbody photons inside a box at two temperatures $T_1$ and $T_2$, and mean temperature $T_{\rm b}=\frac{1}{2}(T_1+T_2)$ initially (left panel). Thomson scattering mixes the two photon distributions without changing the photon number or energy. The averaged distribution is not a pure blackbody but at second order in the temperature difference exhibits a $y$-type distortion in the Wien tail (right panel). This initiates the thermalization process in each patch and repeated Compton scattering slowly converts the distortion into a $\mu$-distortion. The figure adapted from \citet{Chluba2015features}.}
   \label{fig:Superposition}
\end{figure}
\subsection{Damping of primordial small-scale perturbations}
\label{sec:damp}
The damping of small-scale fluctuations of the CMB temperature set up by inflation at wavelength $\lambda<1\,\Mpc$  causes another {\it inevitable} distortion of the CMB spectrum \citep{Sunyaev1970diss, Daly1991, Barrow1991}. The idea behind this mechanism is extremely simple and just based on the mixing of blackbodies with varying temperatures through Thomson scattering (see Fig.~\ref{fig:Superposition}). However, the process was only recently described rigorously \citep{Chluba2012, Khatri2012short2x2, Pajer2012b, Inogamov2015}, allowing us to perform detailed computations of the associated distortion signal for different early-universe models. 

\begin{figure} 
   \centering
   \includegraphics[width=0.87\columnwidth]{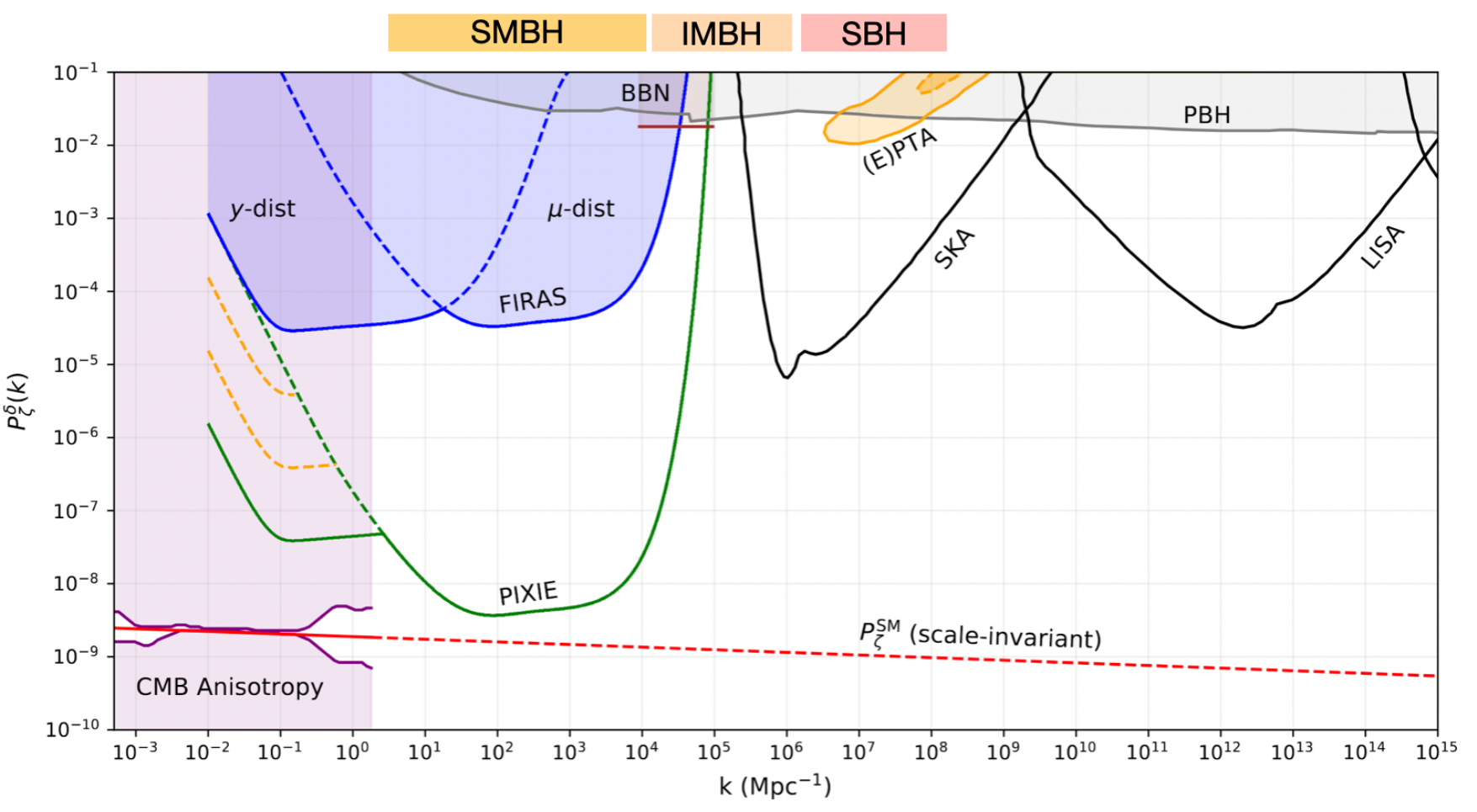}
   \caption{Current constraints on the small-scale power spectrum. At large scales ($k\lesssim 3\,{\rm Mpc}^{-1}$), CMB anisotropies and large scale structure measurements provide very stringent limits on the amplitude and shape of the primordial power spectrum. At smaller scales, the situation is much more uncertain, and at $3\,{\rm Mpc}^{-1}\lesssim k\lesssim 10^4\,{\rm Mpc}^{-1}$, which can be targeted with CMB spectral distortion measurements, wiggle room of at several orders of magnitude is still present. CMB distortion measurements could improve these limits to a level similar to the large-scale constraints. At still larger scales, gravitational wave experiments can set constraints by non-detection of a stochastic background induced through scalar induced gravitational waves (the PTA, SKA, and LISA contours). In addition, large scalar fluctuations can seed abundant PBHs by horizon-scale collapse, which is constrained by various astrophysical and cosmological signatures \citep[e.g.,][]{Carr2010}. We indicated the scales for primordial supermassive BH (SMBH), intermediate mass BH (IMBH) and stellar BH (SBH) formation, which CMB spectral distortions can shed light on directly (i.e., SMBH \& IMBH formation), or indirectly through enhanced $\mu$-distortions for primordial SBHs. The figure is adapted from \cite{Cyr2024PBH}.}
   \vspace{-4mm}
   \label{fig:constraints}
\end{figure}

The distortion is sensitive to the amplitude and shape of the power spectrum at very small scales (corresponding to wavenumbers $1\,\Mpc^{-1}\lesssim k \lesssim \pot{2}{4}\,\Mpc^{-1}$ corresponding to multipoles $10^5\lesssim \ell \lesssim 10^8$) and thus provides a promising new way for constraining inflation, while the relevant modes are still evolving linearly \citep{Chluba2012inflaton}. 
In the early days of CMB cosmology, this effect was already used to derive the first upper limits on the spectral index of scalar perturbations, yielding $n_{\rm S}\lesssim 1.6$ from \COBEF \citep{Hu1994}. Perturbation modes with $1\,\Mpc^{-1}\lesssim k \lesssim 50\,\Mpc^{-1}$ create $y$-distortions, while modes with $50\,\Mpc^{-1}\lesssim k \lesssim \pot{2}{4}\,\Mpc^{-1}$ yield $\mu$-distortions. These scales are hard to access by any other means but spectral distortions provide a new sensitive probe in this regime, which could shed light on primordial black hole formation processes (see Fig.~\ref{fig:constraints}).

For a given initial power spectrum of perturbations, the effective heating rate in general has to be computed numerically. However, at high redshifts the tight coupling approximation for the photon transfer functions \citep{Hu1995CMBanalytic} can be used to simplify the calculation. An excellent approximation for the effective heating rate can be obtained using\footnote{Here, we define the heating rate such that $\int_z^\infty \frac{\id (Q/\rho_\gamma)}{\id z}\id z>0$.}
\begin{align}
\label{eq:adiabatic_damping}
\frac{\id (Q/\rho_\gamma)}{\id z}&\approx 4 A^2 \partial_z \kD^{-2} \int^\infty_{k_{\rm min}} \frac{k^4\id k}{2\pi^2} P_\zeta(k)\,\expf{-2k^2/\kD^2},
\end{align}
where $P_\zeta(k)=2\pi^2\,A_{\rm s}\,k^{-3}\,(k/k_0)^{\nS-1 + \frac{1}{2}\,\nrun \ln(k/k_0)}$ defines the usual curvature power spectrum of scalar perturbations, with amplitude $A_{\rm s}$ at pivot scale $k_0$, spectral index $\nS$, running $\nrun$; Here, $\kD$ is the photon damping scale \citep{Weinberg1971, Kaiser1983}, which early on scales as $\kD\approx \pot{4.048}{-6}\,(1 + z)^{3/2} \Mpc^{-1}$. For adiabatic modes, we obtain a heating efficiency $A^2\approx (1+4R_\nu/15)^{-2}\approx 0.813$, where $R_\nu\approx 0.409$ for $N_{\rm eff}=3.046$. The $k$-space integral is truncated at $k_{\rm min}\approx 0.12\,\Mpc^{-1}$, which reproduces the full heating rate across the recombination era quite well. With this formula, we can directly compute the associated distortion using {\tt CosmoTherm}. The various isocurvature perturbations can be treated in a similar manner \citep{Chluba2013iso}; however, for the standard inflation model these should be small. Tensor perturbations also directly contribute to the dissipation process, but the associated heating rate is orders of magnitudes lower than for adiabatic modes even for very blue tensor power spectra \citep{Ota2014, Chluba2015, Kite2021}.

For $A_{\rm s}=\pot{2.207}{-9}$, $\nS=0.9645$ and $\nrun=0$ \citep{Planck2015params}, we present the result in Fig.~\ref{fig:signals}. The adiabatic cooling distortion (see Sect.~\ref{sec:ad_cool}) was simultaneously included. The signal is uncertain to within $\simeq 10\%$ in $\Lambda$CDM, simply because of the remaining uncertainties in the measurement of $A_{\rm s}$ and $\nS$. It is described by a sum of $\mu$- and $y$-distortion with $\mu\approx \pot{2.0}{-8}$ and $y\approx \pot{3.6}{-9}$ and a non-vanishing overall residual at the level of $\simeq 20\%-30\%$. 
A non-detection of this signal would be a direct disproof of $\Lambda$CDM, and possibly slow-roll inflation, no matter what!

In terms of raw sensitivity, the small-scale damping signal is close to the detection limit of a \PIXIE-like experiment; however, foregrounds in particular at low frequencies render a detection more challenging \citep{abitbol_pixie}. Still, a \PIXIE-like experiment could place interesting constraints on the amplitude of scalar fluctuations around $k\simeq 10^3\,\Mpc^{-1}$ \citep[e.g.,][]{Chluba2012inflaton}, potentially helping to shed light on the small-scale crisis \citep{Nakama2017} and rule out models of inflation with increased small-scale power \citep{Clesse2014, Cyr2024PBH}. This exciting perspective has made a future CMB spectrometer one of the strong contenders for an ESA L-class mission in the category "Probes of the early Universe" in the Voyage 2050 program.\footnote{\url{https://www.esa.int/Science_Exploration/Space_Science/Voyage_2050_sets_sail_ESA_chooses_future_science_mission_themes}}

\subsection{Adiabatic cooling for baryons}
\label{sec:ad_cool}
The adiabatic cooling of ordinary matter continuously extracts energy from the CMB photon bath by Compton scattering, leading to another small but guaranteed distortion that directly depends on the baryon density and helium abundance. The distortion is characterized by {\it negative} $\mu$- and $y$-parameters at the level of $\simeq \pot{\rm few}{-9}$ \citep{Chluba2005, Chluba2011therm, Khatri2011BE}. The effective energy extraction history is given by
\begin{align}
\label{eq:adiabatic_cooling}
\frac{\id (Q/\rho_\gamma)}{\id z}\!=\!-\frac{3}{2}\,\frac{N_{\rm tot} k\TCMB}{\rho_\gamma (1+z)} 
\!\approx\! -\frac{\pot{5.7}{-10}}{(1+z)}\,
\!\left[\frac{(1-\Yp)}{0.75}\right]
\!\left[\frac{\Omega_{\rm b}h^2}{0.022}\right]
\left[\frac{(1+f_{\rm He}+X_{\rm e})}{2.25}\right]\left[\frac{T_0}{2.726\,\Kel}\right]^{-3}
\end{align}
where $N_{\rm tot}=N_{\rm H}(1+f_{\rm He}+X_{\rm e})$ is the number density of all thermally coupled baryons and electrons; $N_{\rm H}\approx \pot{1.881}{-6}\,(1+z)^3\,\cm^{-3}$ is the number density of hydrogen nuclei; $f_{\rm He}\approx \Yp/4(1-\Yp)\approx 0.0819$ [using $\Yp=0.25$] and $X_{\rm e}=\Ne/N_{\rm H}$ is the free electron fraction, which can be computed accurately with {\tt CosmoRec}. For {\it Planck} 2015 parameters, the signal is shown in Fig.~\ref{fig:signals}. It is uncertain at the $\simeq 1\%$ level in $\Lambda$CDM and cancels part of the expected damping signal; however, it is roughly one order of magnitude weaker and cannot be separated at the currently expected level of sensitivity of next generation CMB spectrometers, although it can be accurately predicted.

\begin{figure} 
   \centering
   \includegraphics[width=0.88\columnwidth]{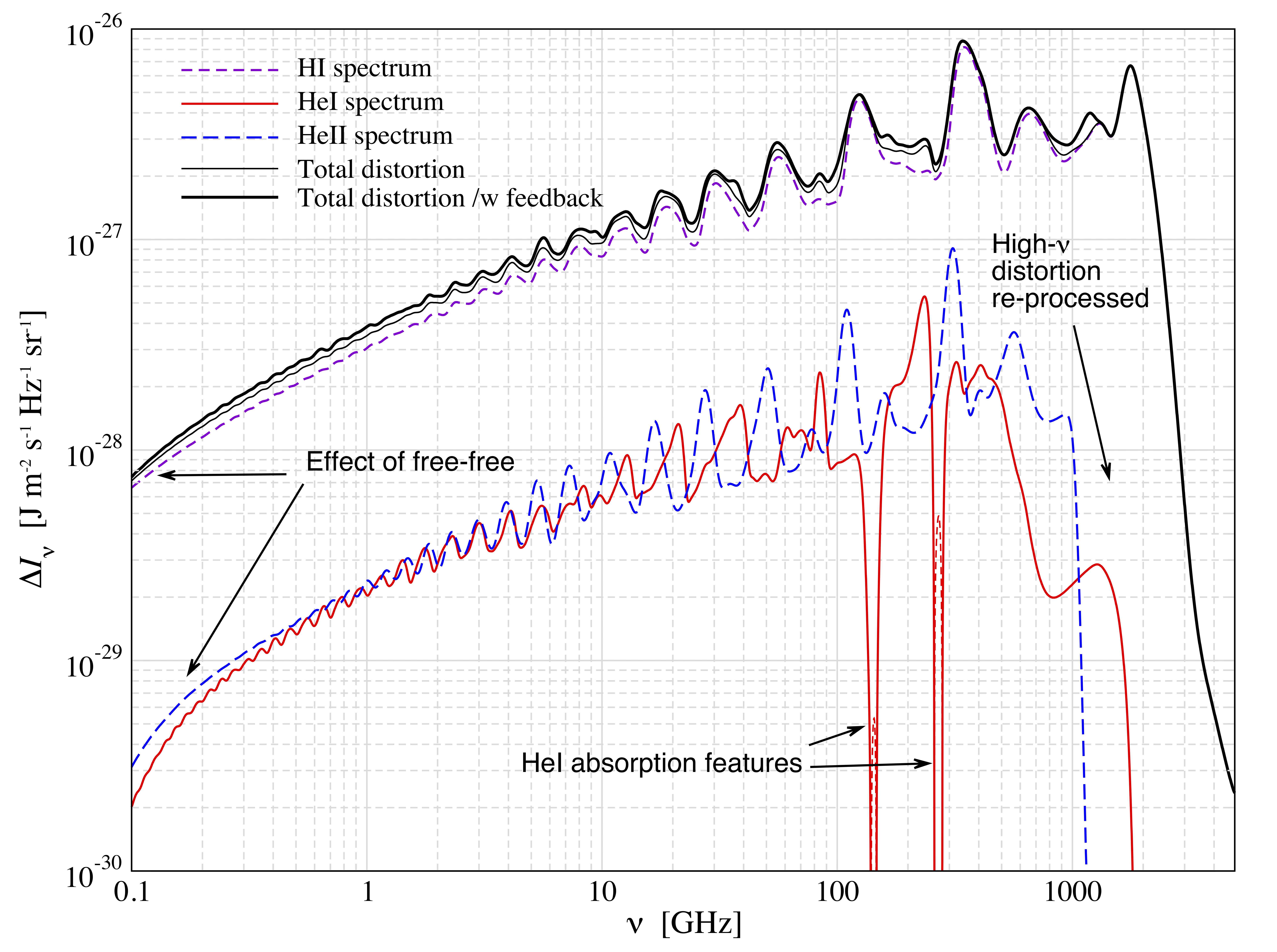}
   \caption{Cosmological recombination radiation from hydrogen and helium for calculations with 500 shell atoms ($\simeq 125,000$ atomic levels per species). The different curves show individual contributions (without feedback) as well as the total distortion with and without feedback processes. At low frequencies, free-free absorption becomes noticeable. The effect is stronger for the contributions from helium due to the larger free-free optical depth before recombination ends at $z\simeq 10^3$. In total, some $6.1\gamma$ are emitted per hydrogen atom when all emission and feedback are included. Hydrogen alone contributes about $5.4\gamma /N_{\rm H}$ and helium $\simeq 0.7\gamma /N_{\rm H}$ ($\simeq 8.9 \gamma /N_{\rm He}$). The Figure was taken from \citet{Chluba2016CosmoSpec}.} 
   \label{fig:recombination}
\end{figure}
\subsection{The cosmological recombination radiation}
The cosmological recombination process is associated with the emission of photons in free-bound and bound-bound transitions of hydrogen and helium \citep{Zeldovich68, Peebles68, Dubrovich1975}. This causes a small distortion of the CMB and the redshifted recombination photons should still be visible as the cosmological recombination radiation (CRR), a tiny spectral distortion ($\simeq$ nK-$\mu$K level) present at mm to dm wavelength \citep[for overview see][]{Sunyaev2009}. The amplitude of the CRR depends directly on the number density of baryons in the Universe. The helium abundance furthermore affects the detailed shape of the recombination lines, while the number of neutrinos has a minor effect. Finally, the line positions and widths depend on when and how fast the Universe recombined. The CRR thus provides an independent way to constrain cosmological parameters and  map the recombination history. 

Several computations of this CRR have been carried out in the past \citep[see][for references]{Sunyaev2009}. These calculations were very time-consuming, taking a few days of supercomputer time for one cosmology. The computational challenge was overcome using spectral conductances for hydrogen and helium \citep{Yacine2013RecSpec, Chluba2016CosmoSpec}, allowing us to compute the CRR in about 15 seconds on a standard laptop using {\tt CosmoSpec}. 
The {\it fingerprint} from the recombination era shows several distinct spectral features that encode valuable information about the recombination process (Fig.~\ref{fig:signals}). Many subtle radiative transfer and atomic physics processes can now be included by {\tt CosmoSpec}, yielding the most detailed and accurate predictions of the CRR in the standard $\Lambda$CDM model to date (see Fig.~\ref{fig:recombination}). In $\Lambda$CDM, the CRR is uncertain at the level of a few percent, with the error being dominated by atomic physics rather than cosmological parameter values.

The CRR is several times below the level of the $\Lambda$CDM $\mu$-distortion, and a detection from space will require exquisite sensitivity \citep{Vince2015, Hart2020CRR}, which might come into reach of a dedicated CMB spectrometer in the Voyage 2050 program \citep{Chluba2021Voyage}. At low frequencies ($1\,\GHz\lesssim \nu\lesssim 10\,\GHz$), the significant spectral variability of the CRR may also allow us to detect it from the ground with APSERa \citep{Mayuri2015}.
This could open a new way for directly studying the conditions of the Universe at $z\simeq 10^3$ (hydrogen recombination), $z\simeq 2000$ (neutral helium recombination) and $z\simeq  6000$ (singly-ionized helium recombination). 
Furthermore, if something unexpected happened during the recombination stages, atomic species will react and produce additional distortion features that can exceed those of the normal recombination process \citep{Chluba2008c, Chluba2010a}.

To appreciate the importance of the cosmological recombination process at $z\simeq 10^3$ a little more, consider that today measurements of the CMB anisotropies are sensitive to uncertainties of the ionization history at a level of $\simeq 0.1\%-1\%$ \citep{Fendt2009, Jose2010, Shaw2011}. For a precise interpretation of CMB data, uncertainties present in the original recombination calculations had to be reduced by including several previously omitted atomic physics and radiative transfer effects. This led to the development of the new recombination modules {\tt CosmoRec} and {\tt HyRec}, which are now used in the analysis of CMB data. Without these improved treatments of the recombination process the value for $\nS$ would be biased by about $3\sigma$ \citep{Shaw2011, Planck2015params}. We would be discussing different inflation models without these corrections taken into account! Conversely, this emphasizes how important it is to {\it experimentally} confirm the recombination process. 
In light of the {\it Hubble tension}, the CRR might become {\it the} crucial test of our theoretical assumptions that no other probe can deliver \citep{Hart2023CRRHubble, Lucca2024CRR, Lynch2024DESI}.

\begin{figure} 
   \centering
   \includegraphics[width=0.9\columnwidth]{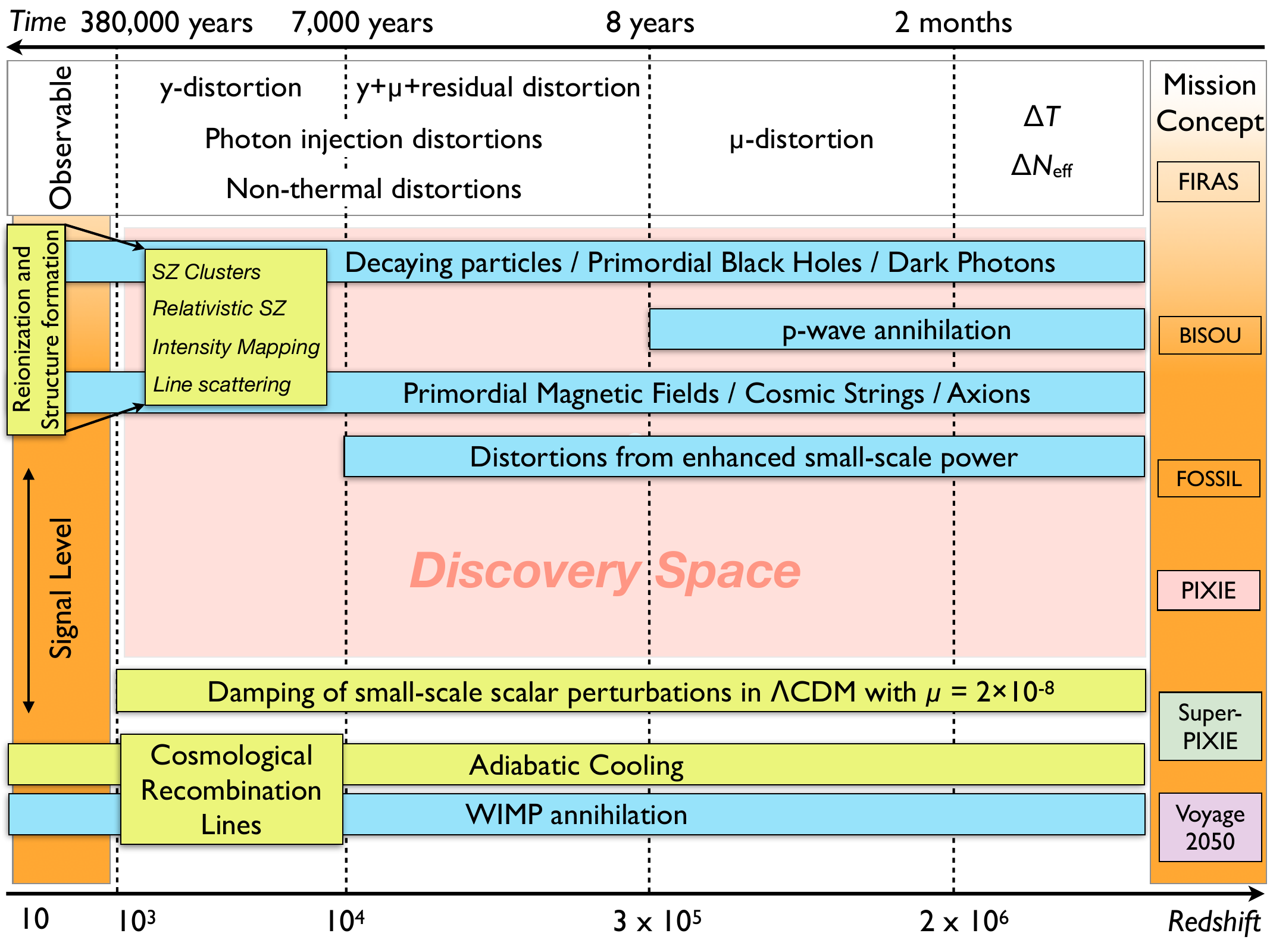}
   \caption{Science thresholds and mission concepts of increasing sensitivity. Guaranteed sources of distortions and their expected signal levels are shown in yellow), while non-standard processes with possible signal levels are presented in turquoise. Spectral distortions could open a new window to the pre-recombination Universe with a vast {\it discovery space} to new physics that is accessible along the path towards a detection and characterization of the $\mu$-distortion from the dissipation of small-scale acoustic modes set by inflation and the cosmological recombination radiation. The figure adapted from \citet{Chluba2021Voyage}.} 
   \label{fig:scenarios}
\end{figure}

\subsection{New physics examples}
In the previous sections, we have highlighted the main $\Lambda$CDM distortions. The signals are {\it guaranteed} and -- in the absence of surprises -- can be predicted with a high level of confidence. This means that we have several distortion targets that we can use to design and optimize our CMB spectrometers for. However, by no means should we forget the vast discovery space that CMB distortions can probe (see Fig.~\ref{fig:scenarios}). CMB SDs can generally shed light on any process that causes changes to the photon-baryon fluid. Direct electromagnetic interactions (e.g., heating or photon injection/removal) are the most obvious channels for creating SDs, but even gravitational interactions that indirectly affect the evolution and sourcing of photon-baryon perturbations are extremely relevant. 

The broad science case including new physics examples was made many times, e.g., with \citet{Chluba2019BAAS} and \citet{Chluba2021Voyage} highlighting some of the opportunities for the NASA and ESA Voyage 2050 space program, which we refer to for references. 
The range of examples includes {\it inflation scenarios}, {\it decaying} or {\it annihilating} (DM) particles, {\it primordial black holes}, {\it primordial magnetic fields}, {\it axions}, {\it dark photons}, {\it neutrino interactions} and many more. Clearly, many of these scenarios are commonly considered by the broad science community in connection with some of the remaining puzzles and possible extensions in cosmology.

Without wanting to going into details about each single one of the aforementioned scenarios and how they in detail affect the CMB spectrum or may be distinguished by combining different probes (all extremely rich topics of research), one should stress strongly that the measurements of \COBEF even 30 years later often still provide the only or at least most robust constraints to models. In reality, in some cases the value of \COBEF can only be fully appreciated now, as detailed computations of the constraints have become possible with modern tools like {\tt CosmoTherm}. It is therefore imperative to continue exploring and refining the treatments of CMB SD physics and also advance CMB spectroscopy after more than 30 years of dormancy. We are entering an exciting stage in CMB SD science with upcoming experiments like BISOU and TMS, which will help in paving the path towards a future CMB spectrometer mission to open an entirely new window to the early Universe.

\subsection{Anisotropic CMB distortions}
To close the discussion of different distortion signals, we briefly mention anisotropic ($\leftrightarrow$ {\it spectro-spatial}) CMB distortions.
Even in the standard $\Lambda$CDM cosmology, anisotropies in the spectrum of the CMB are expected. The largest guaranteed source of SD anisotropies is due to the Sunyaev-Zeldovich effect caused by the hot plasma inside clusters of galaxies \citep{Zeldovich1969, Sunyaev1980, Birkinshaw1999, Carlstrom2002, Mroczkowski2012}, as already mentioned above. The $y$-distortion power spectrum has already been measured directly by \Planck \citep{Planck2013ymap, Planck2016ymap} and encodes valuable information about the atmospheres of clusters \citep[e.g.,][]{Refregier2000, Komatsu2002, Battaglia2010, Dolag2016}. Similarly, the warm hot intergalactic medium contributes and should become visible \citep{Zhang2004, Dolag2016}. 

A second source of primordial SD anisotropies can be caused by anisotropic heating from the damping of acoustic modes. This is because the damping signal is sensitive to primordial non-Gaussianity in the squeezed-limit, usually parametrized using $f_{\rm NL}$, leading to a spatially-varying spectral signal 
that correlates with CMB temperature anisotropies as large angular scales \citep{Pajer2012, Ganc2012}.
In simple words, the amplitude of small-scale fluctuations in a patch in one direction is slightly different from that in another. For Gaussian fluctuations, this is a rather minor effect (arising from patch to patch variance terms), but in the presence of primordial non-Gaussianity, the modulation of the power can be large and is linked to large-scale modes, leading to the aforementioned correlation.
This effect therefore provides a unique way for studying the scale-dependence of $f_{\rm NL}$ \citep{Biagetti2013, Ravenni2017}, with constraints that can be obtained with existing and planned CMB experiments such as \Planck, \Litebird and CMB-S4 \citep{Remazeilles2018, Rotti2022, Zegeye2023}.
CMB spectral distortion anisotropies hence deliver a complementary and independent probe of early-Universe physics, which allows capitalizing on the synergies with large-scale $B$-mode polarization measurements. 
For foreground mitigation, observations with the SKA may enhance the capabilities of CMB imagers by an order of magnitude, bringing the constraints down to levels similar to those reached at large angular scales with CMB anisotropies \citep{Zegeye2024SKA}. These new physics targets could be extremely important for studies of multi-field inflation models, allowing us to test models in regimes that we cannot probe otherwise \citep{Dimastrogiovanni2016, McCulloch2024}.

To close, another extremely exciting opportunity has now emerged which allows us to model the full spectro-spatial evolution of distortions from various scenarios. This has shown that even average energy release in the perturbed universe sources anisotropic SDs that can be used to constrain SD scenarios \citep{Chluba2023FHII, Kite2023FHIII}. In this case, not only the spectrum itself contains information about the physics and epoch-dependence of the processes, but also the (spatial) correlations with the temperature and polarization anisotropies shed light on this. We have only started to explore this new direction of research, which can capitalize strongly on the experience of the cosmology community with studies of the CMB anisotropies.  

\section{Outlook}
Ever increasing data volumes and more accurate cosmological measurements lie ahead, but to gain deeper insight into the fabric of the cosmos we ultimately need to access new observables, such as the mostly unexplored CMB spectral distortions. CMB spectral distortion measurements provide a unique way of studying physical processes leading to energy release or photon injection in the pre- and post-recombination eras. In the future, this could open a new unexplored window to early-universe and particle physics, delivering independent and complementary pieces of information about the Universe we live in. 
We highlighted several processes that should lead to distortions at a level within reach of present-day technology. Different distortion signals can be computed precisely and efficiently for various scenarios using both analytical and numerical schemes.  Epoch-dependent information, beyond the standard $\mu$- and $y$-type parametrization, may allow us to distinguish various physical scenarios. The cosmological recombination radiation will allow us to check our understanding of the recombination processes at redshifts of $z\simeq 10^3$, and possibly play an important role in solving existing tensions in cosmology. 
Novel opportunities with SD anisotropies and their correlations furthermore give unique science targets for existing, ongoing and planned CMB imagers such as \Planck and \Litebird.
All this emphasizes the immense potential of CMB spectroscopy, both in terms of {\it discovery} and {\it characterization} science, and we should make use of this invaluable source of information in the near future.

\vspace{4mm}
\hrule
\hrule
\vspace{-2mm}
\subsection*{Various codes}
\begin{itemize}
\setlength{\itemindent}{\listindentation}
\item {\tt BRpack} was introduced in \citet{Chluba2020BRpack} for precise modeling of Bremsstrahlung

\item {\tt CosmoRec} was introduced in \citet{Chluba2010b} for precise computations of the cosmological recombination problem

\item {\tt CosmoSpec} was introduced in \citet{Chluba2016CosmoSpec} for computations of the cosmological recombination radiation

\item {\tt CosmoTherm} was introduced in \citet{Chluba2011therm} for the accurate modeling of CMB spectral distortion scenarios

\item {\tt DCpack} was introduced in \citet{Ravenni2020DC} for precise modeling of double Compton emission

\item {\tt HyRec} was introduced in \citet{Yacine2010c} for precise computations of the cosmological recombination problem

\item {\tt SZpack} was introduced in \citet{Chluba2012SZpack} for modeling the various SZ effects and their relativistic corrections

\end{itemize}
\vspace{3mm}
\hrule
\hrule

%

\bibliographystyle{Harvard}
\bibliography{els-article, further}

\end{document}